\documentclass[letterpaper]{article} 
\usepackage{aaai25}  
\usepackage{times}  
\usepackage{helvet}  
\usepackage{courier}  
\usepackage[hyphens]{url}  
\usepackage{graphicx} 
\usepackage{natbib}  
\usepackage{caption} 
\usepackage{algorithm}
\usepackage{algorithmic}

\usepackage[utf8]{inputenc}
\usepackage{amssymb}  
\DeclareUnicodeCharacter{0308}{}  
\DeclareUnicodeCharacter{2260}{\neq}  

\usepackage{newfloat}
\usepackage{listings}
\DeclareCaptionStyle{ruled}{labelfont=normalfont,labelsep=colon,strut=off} 
\floatstyle{ruled}
\newfloat{listing}{tb}{lst}{}
\floatname{listing}{Listing}
\usepackage{float}
\usepackage{tcolorbox}
\tcbuselibrary{skins,breakable}
\usepackage{booktabs}
\usepackage{array}
\usepackage{mdframed}
\usepackage{xcolor}

\title{PETLP: A Privacy-by-Design Pipeline for Social Media Data in AI Research}

\author {
    Nick Oh\textsuperscript{\rm 1},
    Giorgos D. Vrakas\textsuperscript{\rm 2},
    Si\^an J.M. Brooke\textsuperscript{\rm 3,4},
    Sasha Morini\`ere\textsuperscript{\rm 5}\textsuperscript{\rm †},
    Toju Duke\textsuperscript{\rm 6}
}
\affiliations {
    \textsuperscript{\rm 1}socius labs\\
    \textsuperscript{\rm 2}University of Cyprus\\
    \textsuperscript{\rm 3}University of Amsterdam\\
    \textsuperscript{\rm 4}London School of Economics and Political Science\\
    \textsuperscript{\rm 5}Conspiracy Watch\\
    \textsuperscript{\rm 6}Bedrock AI\\
    nick.sh.oh@socius.org, vrakas.giorgos-dimitris@ucy.ac.cy, s.j.m.brooke@uva.nl, toju@bedrockai.org
}

\begin{document}

\frenchspacing
\maketitle

{\renewcommand{\thefootnote}{}\footnotetext{\textsuperscript{\rm †}Work conducted during affiliation with the Open Data Institute}\addtocounter{footnote}{0}}

\begin{abstract}
Social media data presents AI researchers with overlapping obligations under the GDPR, copyright law, and platform terms -- yet existing frameworks fail to integrate these regulatory domains, leaving researchers without unified guidance. We introduce \textbf{PETLP} (\textbf{P}rivacy-by-design \textbf{E}xtract, \textbf{T}ransform, \textbf{L}oad, and \textbf{P}resent), a compliance framework that embeds legal safeguards directly into extended ETL pipelines. Central to PETLP is treating Data Protection Impact Assessments as living documents that evolve from pre-registration through dissemination. Through systematic Reddit analysis, we demonstrate how extraction rights fundamentally differ between qualifying research organisations (who can invoke DSM Article 3 to override platform restrictions) and commercial entities (bound by terms of service), whilst GDPR obligations apply universally. We demonstrate why true anonymisation remains unachievable for social media data and expose the legal gap between permitted dataset creation and uncertain model distribution. By structuring compliance decisions into practical workflows and simplifying institutional data management plans, PETLP enables researchers to navigate regulatory complexity with confidence, bridging the gap between legal requirements and research practice.
\end{abstract}

\begin{links}
    \link{AIES-25}{AIES25Link}
\end{links}

\section{Introduction}

Social media platforms have become essential data sources for computational and social science research, enabling investigations into political movements \cite{solovev2022hate, iqbal2022left}, labour inequalities \cite{xu2024tight}, discrimination \cite{nesterov2024contentious} and cultural phenomena \cite{nguyen2023extracting}. These platforms provide datasets of unprecedented scale and immediacy, offering insights into societal trends and human behaviour that surpass traditional methods \cite{bundtzen2023data}. Concurrently, methodological advances -- from BERT-based models \cite{ji2020suicidal, ananthakrishnan2022suicidal} to Large Language Models (LLMs) \cite{deng2023llms, alhamed2024using, vuruma2024utilizing} -- have expanded analytical possibilities, yet with a cost of intensifying ethical and regulatory challenges.

The research landscape presents AI researchers with multiple layers of complexity. At the foundational level, the definition of social media itself continues to evolve -- conventional websites now incorporate social features like user profiles, messaging, and content-sharing to enhance engagement \cite{garcia2023detecting, he2024exploring}, blurring boundaries and creating ambiguity about applicable legal and ethical frameworks \cite{terzis2024law}. Regulatory frameworks add a second layer of uncertainty: the GDPR offers limited guidance on platform-to-researcher data sharing \cite{EDMO2022} while employing broad, often undefined concepts such as `personal data' and `scientific research'.

These definitional and regulatory ambiguities intersect with a third layer -- platform governance tensions. Platforms increasingly invoke privacy concerns to restrict data access \cite{bruns2021after, morten2024researcher}, despite legal precedents establishing no reasonable expectation of privacy for publicly accessible content \cite{Romano2010}. Nevertheless, the ethical landscape presents equally significant challenges, with empirical studies demonstrating that fewer than 10\% of social media research publications address ethical implications beyond securing basic Institutional Review Board (IRB) approval \cite{taylor2018mining, chiauzzi2019digital}. The intersection of these three layers -- evolving platform definitions, regulatory gaps, and increasingly restrictive platform policies -- creates a fragmented compliance landscape where researchers face contradictory obligations that ultimately compromise both research quality and public confidence \cite{EDMO2022}.

To navigate these challenges, we propose the \textit{Privacy-by-design Extract, Transform, Load and Present (PETLP)} framework. Unlike conventional data pipelines that treat compliance as an afterthought, PETLP directly addresses each layer of complexity: it clarifies definitional ambiguities, operationalises vague regulatory concepts via living DPIAs, navigates platform restrictions through legally-grounded alternatives, and elevates ethical practice beyond checkbox compliance. The framework integrates three intersecting legal domains -- data protection regulations, intellectual property (IP) rights, and platform contract law -- transforming regulatory requirements from external constraints into structured technical workflows. PETLP addresses the fundamental question: \textit{How can researchers responsibly access, process, and present social media data and derivative models whilst fulfilling legal obligations throughout the data pipeline?}

\section{Background and Scope}
\label{sec:background}

Social media data presents unique challenges for AI research. Its distinctive characteristics, including mixed public-private boundaries, embedded personal information, and complex ownership structures, necessitate careful legal consideration before collection, processing, or model deployment. Understanding the legal frameworks governing such data is therefore essential for responsible AI research practice.

Our analysis examines three intersecting legal domains that collectively govern AI researchers' use of social media data: data protection, IP, and contract law. We centre this analysis on European Union law, which provides a comprehensive, principle-based governance model \cite{lynskey2015foundations, european2021ethics}. The GDPR's extraterritorial reach -- applying to any processing of EU-located individuals' data regardless of researchers' location -- makes compliance virtually certain for social media AI research, as global platforms inevitably include EU users' content. This broad applicability establishes EU law as the de facto baseline for international research standards.

Whilst focusing on current requirements, we acknowledge the EU AI Act (effective August 2025) introduces complementary development-phase obligations that reinforce transparency and risk management principles (see Appendix C). Researchers outside Europe should consult jurisdiction-specific guidance, such as \citet{brown2024web} for US web scraping or the AI Governance and Regulatory Archive (AGORA) \cite{arnold2024introducing} for cross-jurisdictional analysis. However, GDPR's extraterritorial application means most social media AI research requires European compliance regardless of researcher location, making our framework broadly applicable.

\subsection{Defining Social Media Data}

Social media data refers to content generated and collected from users and their interactions within platforms, often in the context of \textit{social relationships}\footnote{The European Data Protection Board (EDPB) defines social media as `online platforms that enable the development of networks and communities of users, among which information and content is shared', where key characteristics include account creation, content sharing, and user connections \cite[p.~4]{edps2021socialmedia}.} \cite{olteanu2019social, stieglitz2018social, batrinca2015social}. This encompasses traditional social networking sites like Facebook (now Meta) and Twitter (now X), alongside content-sharing platforms (YouTube, Instagram), messaging applications with social features (WhatsApp, WeChat), and community-based platforms (Reddit, Discord) \cite{boyd2007social, obar2015social}.

This social dimension creates unique legal and ethical challenges, as individuals can be identified not merely through their own posts but through their networks of connections, interaction patterns, and community memberships -- even when traditional identifiers are removed. A user's participation in specific communities, commenting patterns, or social graph can serve as digital fingerprints, making anonymisation particularly difficult and raising the stakes for privacy protection \cite{qian2019social, jiang2018sa, tian2018deeply}.

\subsection{Legal Foundations for Social Media Data Pipelines}
\label{sec:legalfoundations}

Utilising social media data for AI training requires researchers to traverse a landscape marked by fundamental legal uncertainty \cite{longpre2024position, dornis2025generative}. Three distinct yet interconnected legal regimes govern this terrain, each imposing specific obligations that shape every stage of the research pipeline. Understanding these foundations is essential -- not merely for compliance, but for designing research that can withstand scrutiny, maintain access to data sources, and preserve public trust.

The GDPR establishes comprehensive privacy principles yet provides limited operational guidance for research contexts. Key concepts remain broadly defined \cite{EDMO2022}: what constitutes `research purposes', how `personal data' extends in social media contexts, or which activities qualify as `scientific research'. Similarly, IP law presents ambiguity -- whilst the EU's Text and Data Mining (TDM) exceptions permit researchers to extract patterns from copyright protected data, it remains unclear whether these protections extend to training AI models. Contract law adds further complexity through platform Terms of Service (ToS) that often prohibit automated access, redistribution, or AI training, creating potential conflicts with statutory research rights.

These overlapping regimes create compound risks. A single research project will typically navigate data subject rights under the GDPR, copyright protections on user-generated content, database rights held by platforms, and contractual restrictions imposed through ToS. The following sections examine each legal domain in detail, establishing the conceptual foundations necessary for understanding their technical implications and practical intersections throughout the research pipeline.

\subsubsection{GDPR}

The GDPR establishes fundamental principles for data handling that directly shape how AI researchers can work with social media data. Its definition of `processing' encompasses virtually every research activity -- collecting posts, storing usernames, cleaning datasets, training models, and publishing results all constitute `processing' under the regulation. This means GDPR compliance is not a one-time checkbox but an ongoing obligation throughout the research pipeline, from initial data collection to final model deployment.

Three overarching principles guide GDPR obligations. First, \textit{necessity and proportionality} require that data processing be both essential for research objectives and balanced against privacy impacts. Researchers cannot scrape entire subreddits \textit{just in case} some posts prove useful -- each data point must serve the stated research purpose. Second, \textit{accountability} demands that researchers not only comply with the GDPR but demonstrate compliance through documentation and safeguards. What constitutes `appropriate' safeguards depends on risk: analysing public product reviews requires basic security measures, while processing mental health discussions demands careful handling. Third, \textit{rights and freedoms} of data subjects remain paramount -- individuals retain rights to access, correct, or object to processing of their data, though Article 89 allows some limitations for scientific research where these rights would `seriously impair' research objectives. Article 5 translates these into six specific principles, and Article 25 mandates that these protections be embedded into research design from the outset -- \textit{privacy by design, not by afterthought}.

The following discussion introduces four core GDPR considerations for social media research: the personal data status of social media content, controllership determination, impact assessment requirements, and legal bases for processing. Appendix A supplements this overview, providing the underlying legal rationale, regulatory interpretations, and citations that substantiate these requirements.

\noindent\textit{Social Media Data as Personal Data} Social media data qualifies as personal data under the GDPR, irrespective of public accessibility. The determining factor lies not in the data's availability but in its potential to identify individuals directly or indirectly \cite{WP29DPIAGuidelines, Schrems2024}. Under the GDPR, data is rendered anonymous \textit{only} when individuals cannot be identified by any means `reasonably likely to be used' (Recital 26), accounting for future technological advances and data linkage possibilities -- a standard rarely achievable with social media's rich behavioural patterns. Moreover, social media data frequently enables inference of special category data under Article 9; recent Court of Justice of the European Union (CJEU) rulings confirm that the mere possibility of inferring protected attributes activates heightened safeguards \textit{regardless of accuracy} \cite{OT2022, Meta2023, Lindenapotheke2023}. This is particularly significant for AI research, where models may inadvertently memorise training data or develop unforeseen inference capabilities \cite{EDPB2024Opinion28}, demanding researchers account for both intended outcomes and latent model capacities.

\noindent\textit{The Researcher as Joint Controller} While research institutions formally act as controllers, controllership is functionally determined by actual influence over data processing decisions \cite{EDPBControllerProcessor2020}. AI researchers directly determine collection methodologies, transformation techniques, and model training approaches -- decisions integral to defining the \textit{means} of processing under Article 4(7). This paper therefore advocates viewing researchers as joint controllers with their institutions, recognising that joint controllership emerges when parties jointly determine purposes and means through complementary decisions \cite{EDPBControllerProcessor2020}. This perspective ensures privacy considerations permeate the entire research pipeline rather than remaining abstract institutional obligations, creating accountability where data handling decisions are actually made (Appendix D, Figure 2).

\noindent\textit{Data Protection by Design and Impact Assessments} For AI researchers using social media data, DPIAs are effectively mandatory under Article 35, as such research routinely triggers multiple high-risk criteria identified by the \citet{WP29DPIAGuidelines}: large-scale processing, dataset combination, and innovative technology deployment (Appendix B). A compliant DPIA must: (1) systematically describe the processing including data flows and infrastructure (Article 35(7)(a)); (2) assess necessity and proportionality while demonstrating Article 5 principle implementation (35(7)(b)); and (3) identify risks with appropriate mitigations (35(7)(c)). Notably, DPIAs are \textit{living} documents requiring updates throughout iterative AI development cycles \cite[p. 9]{WP29DPIAGuidelines}.

\noindent\textit{Establishing a Legal Basis for Processing} Social media researchers typically rely on either public interest (Article 6(1)(e)) or legitimate interests (Article 6(1)(f)) as their legal basis, with the choice determined by organisational context (Appendix D, Figure 3). Public authorities and universities operating under statutory research mandates can invoke public interest grounds, provided they demonstrate that their research serves a recognised public benefit \cite{Agentsia2024}. In contrast, private sector entities must conduct and document a \textit{Legitimate Interest Assessment} (LIA) to justify their processing activities, as outlined in the \citet{WP29DPIAGuidelines}. Both require distinguishing broader research \textit{interests} from specific processing \textit{purposes}. The GDPR affords research significant flexibilities under Article 89 -- including purpose compatibility, extended retention, and conditional rights limitations -- provided researchers implement appropriate safeguards \cite{edpb2021secondaryuse}. These privileges enable research while maintaining privacy protection through proportionate technical and organisational measures.

\subsubsection{IP and Contract Law}

IP and contract law determine the conditions under which social media data may be lawfully accessed, used and shared. Recent AI advances have intensified tensions in the IP landscape, raising questions about the scope of existing rights and exceptions \cite{oecd2025ipai}. Whilst scraping can implicate various rights (trademarks, trade secrets, publicity, moral rights), we focus on copyright and database rights.

\noindent\textit{Copyright and Database Rights} Under the \textit{InfoSoc Directive} \cite{InfoSocDirective2001}, copyright grants authors exclusive economic rights over reproduction, distribution and public communication of their works. The CJEU has set a low originality threshold -- `the author's own intellectual creation' -- holding in \textit{Infopaq International A/S v Danske Dagblades Forening} (C-5/08) that even eleven-word extracts could qualify if bearing the author's personal stamp, and in \textit{Painer v Standard Verlags GmbH } (C-145/10) that portrait photographs merit protection through creative choices in framing and lighting. This means even modest user-generated content (posts, short videos) may be protected if demonstrating individualised creative input. Simultaneously, the \textit{Database Directive} \cite{Directive96_9} establishes \textit{sui generis} rights where `substantial investment' exists in obtaining, verifying or presenting contents, potentially protecting platforms' aggregated collections of posts and metadata. This dual-layer protection -- individual posts via copyright, aggregated collections via database rights -- creates complex questions about scraping and reusing social media datasets, particularly where data provenance is unclear \cite{oecd2025ipai}. To balance these rights with scientific advancement, the \textit{DSM Directive} \cite{DSMDirective2019} introduced mandatory Text and Data Mining (TDM) exceptions: Article 3 grants research organisations unwaivable TDM rights for scientific research on lawfully accessed content, while Article 4 provides broader TDM rights that rightholders may reserve through machine-readable means such as the Robot Exclusion Protocol (\texttt{robots.txt}) -- a text file that specifies which parts of a website automated crawlers may access.

\noindent\textit{Contractual and Technical Safeguards} Contract law introduces an additional layer of legal obligations, often in the form of platform ToS or API agreements. These contractual terms define the conditions under which users, including researchers, may access and interact with platform data. Violating these terms can expose researchers to contractual liability, including account suspension, loss of data access, or even legal action for breach of contract. Whilst the DSM Directive renders certain contractual restrictions on research unenforceable, many limitations remain valid and binding. Platforms also implement \texttt{robots.txt} to control automated access, which researchers should respect as both a legal requirement (where they constitute Article 4 DSM reservations) and an ethical consideration.

\subsection{Research Scope and Contributions}

With these legal foundations established, this paper makes three core contributions. First, we introduce \textbf{PETLP}, a privacy-by-design framework that fundamentally reconceptualises ETL pipelines for the AI era. Rather than retrofitting compliance checkpoints, PETLP embeds DPIAs as living design tools that guide decisions from project conception through post-deployment. While no framework can accommodate every platform's unique policies and social dynamics, PETLP provides a methodological \textit{template}. It offers general interpretive guidance rather than prescriptive instructions, empowering researchers to make informed decisions within their specific contexts.

Second, we provide operational clarity by synthesising three intersecting legal regimes -- data protection, IP, and platform contract law -- into practical decision trees (Appendix D, Figure 2-10). These trees are designed to resolve common ambiguities: when platform terms override statutory rights, how to qualify for DSM Article 3 protections, which privacy safeguards satisfy the GDPR, and whether models can be legally published.

Third, we demonstrate practical application through a systematic Reddit case study. We selected Reddit for its relatively open data access\footnote{While X charges \$200/month for 10,000 posts, Reddit permits 100 queries per minute under academic guidelines \cite{RedditDataAPIWiki2024, xapi2023rate}.} and diverse community structure, which illuminates varied research ethics challenges. This implementation exemplar shows how PETLP navigates platform-specific complexities, providing a replicable model for platform-specific compliance analyses.

\section{The PETLP Framework}
\label{sec:etlp}

\subsection{Revisiting the ETL Model}

Extract, Transform, Load (ETL) pipeline originated in the 1970s as a structured process for converting raw data into formats suitable for analysis and decision-making. At its core, ETL involves three sequential operations: \textit{extracting} data from various sources, \textit{transforming} it into a usable format, and \textit{loading} it into a storage system for analysis. Given the absence of standardised practices for data collection and documentation in social media research \cite{Bundtzen2023}, this model offers several attractive features: (1) providing a systematic structure for enhancing reproducibility and traceability; (2) well-established in both research and industry, facilitating methodological coherence; and (3) its sequential logic enabling a clear articulation of legal and ethical responsibilities at each stage.

However, applying ETL directly to social media research is not without difficulties. Research workflows are rarely linear. As \citet{markham2012ethical} observe, methodological stages often overlap or iterate unpredictably: researchers may extract and load data prior to transformation (\textit{ELT}), delay processing due to ethical review cycles, or conduct exploratory analyses before finalising dataset structure. 

Despite these limitations, we propose that an adapted ETL framework can serve as a valuable conceptual model for responsible data pipeline. We propose a modified framework -- \textbf{PETLP} -- that extends ETL in two key ways. First, the prefix ``\textbf{P}'' stands for \textit{privacy-by-design}, implemented through the default use of DPIAs initiated prior to data collection and updated across all stages of the research lifecycle. This ex-ante approach enables researchers to identify and mitigate privacy risks before they materialise, rather than retrofitting safeguards after data collection. Notably, while PETLP provides structured checkpoints for privacy assessment, it accommodates the iterative nature of research by treating DPIAs as living documents that evolve with methodological refinements. Second, we introduce a fourth phase, \textit{presentation}, supported by established research stage taxonomies \cite{markham2012ethical} and social media research typologies \cite{AndersenSoderqvist2012} recognising dissemination as an indispensable component of the research lifecycle.

In the next sections, we examine each phase of the PETLP framework in detail:

\begin{itemize}
\item \textbf{Privacy-by-Design}: Operationalises GDPR Article 25 through default DPIAs as living documents that guide decisions throughout the research lifecycle.
\item \textbf{Extract}: Analyses four data acquisition channels (platform-authorised, user-mediated, third-party, self-directed) with availability determined by researcher status, showing how copyright exceptions, platform terms, and GDPR obligations intersect differently for each method.
\item \textbf{Transform}: Addresses the dual challenge of copyright reproductions (every preprocessing creates copies requiring authorisation) and privacy engineering (implementing minimisation, anonymisation attempts, and technical safeguards) during data cleaning and preparation.
\item \textbf{Load}: Establishes secure storage architectures, international transfer compliance, and retention governance for transformed data, creating the infrastructure for compliant data access and querying.
\item \textbf{Present}: Encompasses both AI model training and research dissemination (datasets, papers, models, services), evaluating extraction attack risks, copyright liability, platform restrictions, and the tension between open science mandates and privacy protection.
\end{itemize}

\subsection{Privacy-by-Design}
\label{subsec:privacy}

\textit{Privacy-by-design} -- mandated in Article 25 GDPR and reinforced by accountability obligations in Article 24 -- requires privacy safeguards form part of a project's architecture from conception. PETLP operationalises this through DPIAs as \emph{default design tools} spanning the entire research lifecycle. Researchers initiate DPIAs during pre-registration, update them when processing changes, and use them to guide decisions throughout every pipeline phase (Figure 1; see Appendix E for a comprehensive risk assessment example).

During pre-registration, researchers should consult comprehensive risk frameworks to identify potential privacy challenges early. Given the increasing adoption of LLMs in research contexts, \citet{barbera2025aiprivacy} offers particularly valuable guidance tailored to LLM privacy considerations: privacy and data protection risks with recommended mitigations (pp. 28--42), risk identification methodologies (pp. 48--56), estimation and evaluation protocols (pp. 57--65), control and mitigation strategies (pp. 66--73), residual risk assessment (p. 74), and state-of-the-art tools and benchmarks (as of March 2025) (pp. 97--101). Though LLM-focused, these frameworks provide transferable insights for broader AI applications, enabling researchers to anticipate risks during initial DPIA development rather than discovering them retrospectively.

This approach creates significant efficiencies, particularly for researchers relying on legitimate interests under Article 6(1)(f). As the UK Information Commissioner's Office clarifies, a comprehensive DPIA encompasses all elements of a Legitimate Interest Assessment (LIA) whilst providing deeper risk analysis \cite{ico2023}. The DPIA's structured evaluation of purpose, necessity, and proportionality automatically satisfies the LIA's three-part test \cite{EDPB2024Art6Guidelines}, eliminating redundant assessments while creating the documentary evidence required for ongoing compliance.

For social media research -- which invariably triggers multiple high-risk indicators under EDPB criteria -- this approach delivers five key benefits: (1) satisfies mandatory LIA requirements for legitimate interests processing; (2) fulfils Article 35 obligations for high-risk processing; (3) embeds risk-awareness and data minimisation directly into design decisions; (4) provides auditable compliance records for research projects; and (5) establishes structured mechanisms for refining safeguards as processing evolves.

By anticipating data sharing and replication requirements from inception, researchers can design access controls and anonymisation strategies that balance open science principles with privacy protection. This ensures research outputs satisfy both scholarly standards and regulatory requirements without requiring substantial retrospective modifications -- converting compliance from perceived burden into methodological asset.

\begin{figure*}[!htbp]
\centering
\includegraphics[width=\textwidth]{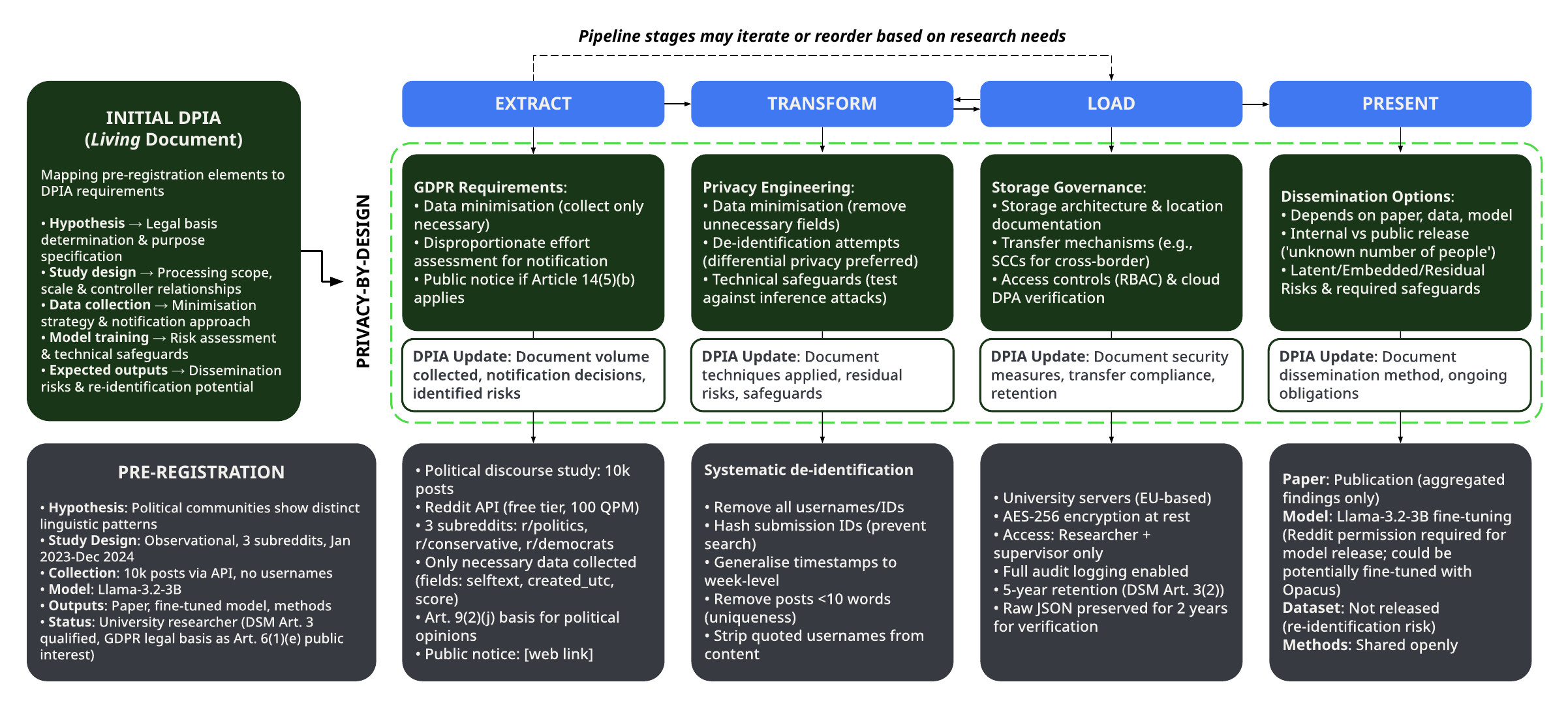}
\caption{\textbf{Privacy-by-design ETLP (PETLP) framework for social media AI research}. \textit{Pipeline stages (Appendix D, Figure 2-10) and Reddit case study (Appendix E)}.}
\end{figure*}

\subsection{Extract}
\label{subsec:extract}

The \textit{Extract} phase addresses the fundamental question of how researchers can lawfully acquire social media data for AI research. Unlike subsequent stages, extraction determines whether research may proceed at all. The legitimacy of data acquisition often serves as the gatekeeping criterion for entire research programmes. 

Recent jurisprudence offers encouraging signals for academic researchers. The Hamburg District Court's decision in \textit{LAION v. Kneschke} (2024) affirmed that creating \texttt{LAION-5B} dataset \cite{schuhmann2022laion} through web scraping constitutes legitimate Text and Data Mining (TDM) for scientific research purposes under the DSM Directive. Notably, the court held that platform terms cannot override statutory research exceptions for qualifying institutions -- a principle codified in DSM Article 7(1).

However, this protection applies only to specific researchers under specific conditions. To navigate these variations, we develop a practice-informed typology of four extraction channels: platform-authorised access, user-mediated collection, third-party aggregation, and self-directed extraction. This typology, derived from regulatory analysis (what the law permits), platform architectures (what is technically feasible), and documented research practice (what researchers actually do), reveals that extraction legality depends on three intersecting factors: the researcher's institutional status (qualifying research organisation or commercial entity), the research purpose (scientific or commercial), and the specific legal basis invoked (DSM Article 3 and 4, and GDPR legal bases) (Appendix D, Figure 6).

\subsubsection{Platform-authorised Access}

Platform-authorised access encompasses official APIs, research partnerships, and developer portals explicitly provided for sanctioned use. This method offers the clearest legal pathway, operating within platform-specific ToS and developer agreements. However, its availability and utility vary significantly by researcher type and platform strategy.

From a contractual perspective, platforms typically impose \textit{browsewrap} agreements -- terms that apply automatically upon site access. Their enforceability depends on adequate notice to users \cite{gupta2012websites, DouglasvTalkAmerica2007}. More robust are click-through agreements required for API access, which create explicit contractual relationships. Reddit exemplifies this tiered approach: whilst content remains publicly viewable, its browsewrap User Agreement prohibits unauthorised scraping (Section 7), effectively channelling researchers toward the official Data API, which requires explicit acceptance of additional Developer and API Terms \cite{RedditUserAgreement, RedditDeveloperTerms, RedditDataAPITerms2024}.

For qualifying research organisations under DSM Article 3, these contractual restrictions may be legally unenforceable when conducting TDM for scientific research (Appendix D, Figure 4 and 5). The \textit{LAION v. Kneschke} decision confirms that non-profit research institutions can invoke this mandatory exception. However, commercial researchers and those outside qualifying institutions remain bound by platform terms, as they must rely on the more limited Article 4 exception, which platforms can contractually exclude.

Despite its legal clarity, platform-authorised access is often ill-suited to the needs of academic research. \citet{bruns2021after} identifies three structural limitations: (1) prohibitively expensive pricing models, (2) restricted export functionalities, and (3) market-oriented data schemas optimised for business analytics rather than open-ended scholarly inquiry. These limitations are not merely logistical; they exert substantive influence over what research questions can be asked and answered. For instance, Reddit's API imposes temporal restrictions that prevent longitudinal analysis beyond six months, fundamentally limiting research designs (Appendix I). In this way, platform-level design choices function as boundary conditions for academic inquiry, shaping not just the data available, but the entire epistemic scope of viable research.

\subsubsection{User-mediated Data Collection}

User-mediated extraction methods rely on active participation by social media users who choose to contribute their data for research purposes. These include \textit{data donations}, participation through browser extensions, or account-linked authorisations that enable researchers to retrieve user data under conditions of explicit consent. Under the GDPR Article 6(1)(a), such explicit consent can serve as a lawful basis for processing, provided it is freely given, specific, informed, and unambiguous. Such methods are often seen as ethically preferable because they foreground transparency, agency, and autonomy. Yet from a legal perspective, user consent does not necessarily override platform-imposed restrictions.

The well-documented case of NYU’s \textit{Ad Observer} project exemplifies the fragility of this method’s legal status. Despite obtaining informed consent from users to collect and share advertising data, the project faced platform resistance. Meta shut down the researchers’ accounts, citing violations of its ToS \cite{Erwin2021}. This signals a regulatory ambiguity -- wherein platforms retain contractual control over data flows, even in the face of user consent.

Pragmatically, this channel is also burdened by logistical and methodological obstacles. Recruiting a sufficiently large and demographically representative user base is often infeasible without major funding or institutional backing \cite{bundtzen2023data}. Studies reliant on data donation or custom-built plug-ins tend to produce biased samples skewed toward technologically literate participants. This raises concerns about the generalisability of findings, particularly for research on platform-wide phenomena.

\subsubsection{Third-party Aggregation Services}

Third-party data aggregators are companies or services that collect social media data on behalf of researchers, eliminating the need for researchers to scrape or access platforms directly. These services -- ranging from commercial vendors like Bright Data to research-oriented platforms like Pushshift (now defunct) -- gather posts, comments, and metadata from social media sites, then provide this pre-collected data to researchers either for free or for a fee. While appealing for their convenience and scalability, these services occupy a precarious legal position.

There are no widely reported cases in the EU that are directly equivalent to \textit{Meta v. Bright Data}, where a court ruled on the third party legality of web scraping public data from social media platforms. However, the EU landscape proves more restrictive. Under the Database Directive Article 7, platforms hold rights over substantial investments in data organisation, prohibiting extraction of substantial parts without authorisation. Article 7(5) extends this to `repeated and systematic extraction' of even insubstantial parts that conflict with normal database exploitation.

Beyond jurisdictional differences, these services face contractual and data protection challenges. Reddit's user agreement, whilst granting the platform broad content rights, explicitly prohibits third-party redistribution without permission \cite{RedditUserAgreement}. The GDPR further complicates matters, requiring aggregators to establish a lawful basis under Article 6 and implement heightened safeguards for special category data under Article 9. 

This suggests that datasets on platforms like Academic Torrents, despite their widespread academic use, exemplify these converging legal risks. Following Reddit's enforcement action against Pushshift, vast archives of Reddit data now circulate via torrents, typically without licenses beyond warnings that content `may be protected by copyright' \cite{AcademicTorrent}. This situation presents researchers with an ethical paradox: datasets fundamental to research exist in legal limbo, potentially violating platform terms, database rights, and data protection law. That these datasets migrated from Pushshift to torrents following platform enforcement underscores their legally dubious status.

\subsubsection{Self-directed Extraction}

Self-directed extraction -- where researchers control their own scraping or automation -- provides the greatest flexibility in data collection but also carries the highest legal risk. Academic and commercial researchers face fundamentally different regulatory landscapes when using this method.

For qualifying research organisations, DSM Article 3 provides robust protection. The \textit{LAION v. Kneschke} court explicitly held that extracting publicly available image URLs for AI dataset constitutes legitimate TDM for research. Platforms may implement measures to ensure the security and integrity of the networks and databases, but these `shall not go beyond what is necessary' (Article 3(3)). Rate limiting might be proportionate; blanket automated access bans likely exceed necessity.

Commercial researchers face a sharply different landscape. Without Article 3 protection, they must rely on Article 4's general TDM exception, which platforms can exclude through machine-readable opt-outs. Reddit's \texttt{robots.txt} and ToS constitute such opt-outs, rendering commercial scraping legally impermissible without prior approval.

Meanwhile, all researchers -- whether academic or commercial -- must recognise that the GDPR operates independently of copyright exceptions. The \textit{LAION v. Kneschke} decision of TDM rights addresses only intellectual property; data protection obligations remain fully applicable. As established in our default DPIA approach, researchers must demonstrate lawful basis, implement data minimisation, and maintain comprehensive documentation throughout the pipeline. The EDPB specifically flags web scraping's risks -- large volume of data collected, the large number of data subjects, and the indiscriminate collection \cite[Paragraph 86]{EDPB2024Opinion28} -- requiring heightened safeguards regardless of TDM authorisation. 

\subsection{Transform}
\label{subsec:transform}

The \textit{Transform} phase focuses on converting raw social media data into analysis-ready formats through cleaning, validation, normalisation, and restructuring operations. Unlike extraction, transformation engages specific legal challenges around intermediate data processing. Each cleaning operation -- removing duplicates, standardising formats, extracting features, or restructuring databases -- creates new copyright reproductions whilst simultaneously constituting the GDPR processing activities. This phase requires researchers to implement DPIA-identified safeguards \textit{and} ensuring their preprocessing activities remain within applicable IP exceptions (Appendix D, Figure 7).

\subsubsection{Data Preprocessing and IP rights}

Data transformation necessarily creates multiple reproductions falling within InfoSoc Article 2's broad scope. Every preprocessing step -- loading data into memory for cleaning, creating intermediate files during normalisation -- constitutes reproduction requiring authorisation. While Article 5(1) InfoSoc exempts `transient or incidental' copies forming an integral and essential part of a technological process, cleaned datasets and extracted features have independent value beyond mere technical facilitation.

For qualifying research organisations, DSM Article 3 permits these reproductions when conducting TDM for scientific research. The \textit{LAION v. Kneschke} decision (Hamburg District Court, 2024) confirms this covers dataset preparation activities, with the court ruling that downloading, analysing, and restructuring data for research purposes falls within the TDM exception. Article 3(2) explicitly protects intermediate datasets, allowing retention of processed copies `for the purposes of scientific research, including for the verification of research results'. 

Commercial researchers, however, face significant constraints. Without Article 3's mandatory protection, they must navigate Article 4's opt-out provisions. Reddit exemplifies comprehensive opt-out strategies, combining technical measures (\texttt{robots.txt}: \texttt{User-agent: *; Disallow: /}), contractual prohibitions \cite[Section 7]{RedditUserAgreement}, and API restrictions \cite{RedditUserAgreement}. The evolution of ``machine-readable'' opt-outs further complicates matters. While the \textit{LAION v. Kneschke} court noted \textit{in obiter} (a non-binding observation that could nonetheless influence future interpretations) that natural language statements might qualify as machine-readable, initiatives like \texttt{llms.txt} \cite{howard2024llmstx} illustrate how machine-readability might evolve beyond traditional interpretations to include human-readable but parser-friendly formats.

\subsubsection{Privacy-Preserving Transformation}

The Transform phase operationalises GDPR principles into concrete preprocessing decisions. Unlike raw data extraction, transformation offers opportunities to embed privacy safeguards. Each decision -- which fields to retain, how to aggregate data, whether to infer missing values -- must balance research utility against privacy principles.

Article 5(1)(c) requires data minimisation: retaining only data `adequate, relevant and limited to what is necessary'. For transformation, this means actively removing unnecessary fields, aggregating where individual-level detail is not required, and resisting the temptation to preserve potentially useful data without clear purpose.

Transformation also provides the primary opportunity for anonymisation -- though achieving true anonymity proves challenging. Removing usernames and IDs (pseudonymisation) offers minimal protection; high-dimensional social media data remains highly re-identifiable. \citet{de2013unique} demonstrated that just four spatial-temporal data points were enough to uniquely identify 95\% of individuals in a 1.5 million-person mobility dataset. Techniques like $k$-anonymity, once standard, have been shown to suffer from both attribute and linkage vulnerabilities, particularly at scale \cite{sweeney2002k, gadotti2024anonymization}.

Against this backdrop, \textit{differential privacy} (DP) has emerged as the state-of-the-art approach for anonymisation, offering provable guarantees against a broad range of adversarial attacks \cite{jiang2021applications}. By adding calibrated noise to aggregated statistics or creating synthetic datasets, DP enables privacy-preserving transformation. Real-world implementations include Social Science One's DP-protected link-sharing statistics on Facebook \cite{nayak2020new}, Wikipedia's pageview metrics \cite{adeleye2023publishing}, and Apple's DP-enabled word detection from Reddit comment histories \cite{hu2023differentially}. However, researchers must recognise that even summary-level data or aggregated features can leak private information. Membership inference, model inversion, and linkage attacks have demonstrated that ``aggregate'' does not equal ``anonymous'' \cite{gadotti2024anonymization}. These vulnerabilities underscore why transformation requires careful consideration beyond simple technical implementation.

Indeed, for research under Article 89, transformation represents the key phase for implementing `appropriate safeguards'. This encompasses both technical measures (encryption, access controls, differential privacy) and \textit{organisational} ones -- documenting transformation logic, maintaining processing logs, and ensuring reproducibility without compromising privacy. Researchers should view transformation not as routine data cleaning but as a proactive opportunity to embed privacy-by-design principles directly into research datasets.

\subsection{Load}
\label{subsec:load}

The \textit{Load} phase migrates transformed social media data into secure storage systems -- databases or cloud repositories -- where it becomes accessible for querying and analysis. Unlike the Presentation phase where model training occurs, Load focuses exclusively on establishing compliant storage infrastructure and \textit{internal governance}. Here, the safeguards defined in the Transform stage are encoded into concrete storage architectures, access protocols, and retention schedules (Appendix D, Figure 8).

\subsubsection{Data Accessibility and Transfer}

Transformed data requires responsible access controls, consistent with the GDPR principles. Under the GDPR, \textit{fully anonymised data} (Recital 26) falls outside its scope. However, the EDPB warns that achieving true anonymisation with social media data proves difficult \cite{WP29DPIAGuidelines}, meaning most stored datasets remain \textit{pseudonymised} and subject to transfer restrictions under Articles 45–49 GDPR: (1) \textit{Adequacy decisions} (Article 45) allow transfers to countries deemed by the European Commission to offer `essentially equivalent' protection to require no additional safeguards; (2) \textit{Appropriate safeguards} (Article 46) apply when transfers to non-adequate countries (e.g., the US, unless covered by the EU-US Data Privacy Framework) require Standard Contractual Clauses (SCCs), Binding Corporate Rules (BCRs), or administrative arrangements; and (3) \textit{Derogations} (Article 49) permit, in exceptional circumstances, explicit consent or public interest justifications to be invoked -- but only for non-repetitive, non-systematic transfers.

These obligations apply when loading data into storage systems across borders -- common in distributed research collaborations. The DPIA must document transfer mechanisms as fundamental steps to demonstrate that personal data is not processed unlawfully \cite[Paragraph 56]{EDPB2024Opinion28}, particularly when using cloud storage providers with global infrastructure.

\subsubsection{Storage Architecture and Security Implementation} 

Security obligations arise immediately upon data storage. Loading data into persistent storage triggers Article 32's mandate for security measures `appropriate to the risk'. The EDPB emphasises that storage systems must implement state-of-the-art protections regardless of complexity or cost \cite[Paragraph 7]{EDPB2024ChatGPTReport}. Essential measures include, but not limited to, role-based access controls limiting data to authorised researchers, encryption both at rest (e.g., Advanced Encryption Standard AES-256) and in transit (e.g., Transport Layer Security TLS 1.3).

Cloud storage requires additional contractual safeguards. Major cloud providers offer pre-approved Data Processing Addenda meeting Article 28 requirements \cite{AWSDPA_2023, GoogleCloudDPA_2025, MicrosoftDPA_2025, SupabaseDPA_2025}. When loading data into these platforms, researchers should verify agreements cover their specific storage architecture. 

\subsubsection{Retention Scheduling and Deletion Protocols} 

Loading data into persistent storage triggers the storage limitation principle under Article 5(1)(e) GDPR, which requires personal data be kept no longer than necessary for the specified purposes. Whilst the GDPR permits extended retention for scientific research under Article 89(1) safeguards, \textit{and} DSM Article 3(2) allows qualifying research organisations to retain data `for the purposes of scientific research, including for the verification of research results', these provisions require operationalisation through concrete retention policies.

Researchers should implement: (1) defined retention schedules specifying maximum storage periods per data category (e.g., 5 years for processed datasets, 2 years for raw API responses); (2) automated deletion triggers using database management tools to enforce these schedules; (3) audit trails documenting retention decisions and deletions; and (4) exception handling procedures for data subject to legal holds or verification requirements.

\subsection{Present}
\label{subsec:present}

The \textit{Present} stage marks the transition from internal governance to public dissemination. It is the point at which insights, outputs, or models derived from social media data are shared with external audiences -- whether through academic publications, public datasets, or open-source distribution of trained models. Whilst earlier stages of the PETLP framework emphasised data acquisition and internal processing controls, this final phase introduces distinct legal and ethical challenges associated with \textit{external exposure}.

The risks introduced at this stage cannot be retroactively addressed by upstream compliance efforts alone. This section examines how researchers must navigate the competing demands of privacy protection when sharing findings, open science mandates requiring data accessibility, and copyright restrictions limiting redistribution -- challenges that emerge uniquely at the point of public dissemination (Appendix D, Figure 9 and 10).

\subsubsection{Data Distribution}

Disseminating research findings requires balancing scientific transparency against re-identification risks and platform redistribution restrictions.

\noindent\textit{Re-identification Risks} Disseminating research findings requires careful alignment with the GDPR's core principles -- especially purpose limitation (Article 5(1)(b)) and data protection by design (Article 25). The Present stage focuses on how outputs may indirectly re-expose personal data, including through seemingly benign presentation methods.

For data distribution, a primary concern is the citation of verbatim content from social media posts. \citet{adams2022scraping} demonstrates that such quotations -- even when stripped of usernames -- enable re-identification through search engines. This risk is especially pronounced in niche subreddits or sensitive thematic contexts, where content is technically public but effectively traceable.

To mitigate this, researchers should calibrate their dissemination methods based on the content's identifiability and subject sensitivity. Strategies may include \textit{paraphrasing} user content instead of quoting verbatim, \textit{aggregation} of findings to group-level insights, \textit{synthetic illustration} using fabricated yet plausible data to demonstrate patterns, and \textit{visualisation methods} that omit individual-level markers.

These decisions must be pre-embedded in the project's DPIA and justified under Article 89's `appropriate safeguards'. Documentation should explicitly address how publication methods prevent the re-identification risks and surveillance effects identified by the EDPB.

\noindent\textit{Transparency-Restriction Tensions} Public presentation requires navigating tensions between transparency mandates and platform restrictions. Reddit's Developer Terms prohibit third-party data redistribution \cite[Section 7.4]{RedditDeveloperTerms}, whilst major funders like Horizon Europe mandate open access to underlying data \cite{eu_guide_open_science_2024}.

Although the \textit{LAION v. Kneschke} decision affirmed that TDM copyright exceptions cover dataset \textit{creation}, it did not extend this protection to \textit{distribution}. The court's narrow focus on reproduction rights leaves researchers without clear legal grounds to redistribute extracted content, even when initial extraction was lawful under DSM Article 3.

For researchers, practical approaches to balance openness with compliance include: (1) \textit{hydration via post IDs}, publishing only content identifiers, allowing reconstruction within platform policies (e.g., RetweetBERT \cite{jiang2023retweet}); (2) \textit{synthetic datasets}, creating statistically equivalent data without redistributing actual posts (e.g., SynthPAI \cite{yukhymenko2024synthetic}); (3) \textit{secure analysis environments}, providing controlled remote access rather than data downloads (e.g., SANE \cite{surf_sane}); and (4) \textit{platform programmes}, leveraging official research access channels (e.g., \texttt{reddit4researchers} \cite{Reddit2024}).

Researchers should document their dissemination strategy within the DPIA, specifying how it balances legal compliance, data minimisation, and research transparency.

\subsubsection{AI model Distribution}

Publishing AI models introduces persistent risks as models may retain personal information, reproduce copyrighted content, and violate platform contractual obligations.

\noindent\textit{Embedded Privacy Risks} The EDPB warns that `AI models trained on personal data cannot, in all cases, be considered anonymous' \cite[Paragraph 34]{EDPB2024Opinion28}, meaning that model publication may constitute ongoing personal data processing. This concern is reinforced by the observation that `information from the training dataset, including personal data, may still remain \textit{absorbed} in the parameters of the model' \cite[Paragraph 31]{EDPB2024Opinion28}. Together, these warnings highlight that model release carries inherent privacy risks that persist beyond the training phase.

While differential privacy techniques such as Differentially Private Stochastic Gradient Descent (DP-SGD) \cite{abadi2016deep}, implemented through tools like \texttt{Opacus} \cite{aketis2025caling}, offer theoretical protection against membership inference, their practical deployment varies significantly by model scale. For smaller models (\textless 100M parameters) typical of classification tasks, \texttt{Opacus} provides effective privacy protection with minimal utility loss. Medium-scale models like BERT (100M-1B parameters) maintain acceptable performance, particularly when fine-tuning rather than training from scratch. Recent advances including FSDP support, and LoRA for parameter-efficient fine-tuning now enable scaling to LLMs like \texttt{Llama-3-8b}. However, \citet{aketis2025caling} focus solely on demonstrating feasibility and computational efficiency for large models, leaving the critical question of model quality and utility under privacy constraints unexplored.

\noindent\textit{Copyright Liability} Model publication introduces copyright risks beyond privacy concerns. We re-emphasise that the \textit{LAION v. Kneschke} decision distinguished dataset creation from model training, explicitly declining to rule whether training and distributing AI models falls within TDM exceptions. This judicial restraint leaves researchers uncertain about model publication rights, even when dataset creation was lawful. Until the pending CJEU case (\textit{Like Company v. Google Ireland}, C-250/25) provides clarity, researchers face three liability scenarios \cite{rosati2024infringing}: (1) \textit{direct liability} when published models generate outputs substantially reproducing training content; (2) \textit{secondary liability} when third parties use published models to create infringing content; and (3) \textit{distribution liability} for sharing datasets containing copyrighted material. Importantly, contractual disclaimers accompanying research outputs provide limited protection. While researchers might attempt to exclude their liability through such provisions (e.g., standardised open-source licences to specify liability limitations), these contractual arrangements typically cannot override statutory copyright protections \cite{rosati2024infringing}. Legal liability may persist despite explicit disclaimers.

\noindent\textit{Contractual Constraints} For social media research, platform-specific restrictions compound these risks. Reddit prohibits using its data for AI training without explicit permission \cite[Section 4.2]{RedditDeveloperTerms}. Although qualifying research organisations may invoke DSM Article 3 for dataset creation, the uncertainty around model training and publication means researchers should implement technical measures to prevent models from reproducing training data verbatim, or consider publishing model architectures and training procedures rather than trained weights.

\section{Discussion}
\label{sec:discussion}

While PETLP successfully navigates overlapping legal regimes as validated through our Reddit case study, it cannot resolve the structural tensions inherent in social media AI research governance.

\textbf{From Checkbox to Continuous Compliance}: Traditional AI governance artifacts -- Model Cards, Data Cards, Transparency Notes -- often deliver only ``checkbox compliance'', satisfying regulatory requirements while failing to enable meaningful oversight \citep{kawakami2024responsible}. PETLP's living DPIA approach embeds compliance decisions directly into research workflows, shifting from performative to operational transparency. This treats privacy as a design principle shaping methodological choices from extraction through deployment. However, even this approach operates within an ecosystem where transparency often legitimises rather than scrutinises AI systems.

\textbf{Navigating Fragmented Governance}: The regulatory landscape lacks coherence, with governance instruments emerging daily from disparate actors \citep{arnold2024introducing}. PETLP structures this complexity through decision trees, yet cannot eliminate underlying fragmentation. The \texttt{LAION-5B} case illustrates a related challenge. Even when datasets satisfy one legal domain (DSM Article 3), they may violate others (the dataset's deprecation following discovery of illegal content \citep{thiel2023generative, NeurIPS_DeprecatedDatasets_2025}). Researchers must therefore navigate not just fragmented governance but also the gaps between legal domains -- where copyright compliance, platform terms, content regulations, and data protection operate independently.

\textbf{Platform Heterogeneity as Structural Barrier}: While \textit{LAION v. Kneschke} protects academic scraping legally, technical enforcement varies unpredictably. Reddit's six-month API limitation constrains longitudinal research; X's pricing excludes unfunded researchers; Meta's account suspensions create chilling effects. PETLP acknowledges this through four extraction channels with distinct trade-offs, yet each platform demands bespoke implementation, multiplying complexity for multi-platform studies.

\textbf{Operationalising Responsible Practice}: PETLP requires complementary infrastructure for practical deployment. First, institutional support through template DPIAs and integrated legal-ethical review would reduce researcher burden. Universities could extend IRB processes to encompass DPIA assessment. Second, shared computational environments with pre-configured differential privacy would enable wider adoption of privacy protections, addressing gaps between theoretical safeguards and practical deployment given documented PII leakage \citep{nasr2023scalable}. While infrastructures like European Open Science Cloud (EOSC) demonstrate federated research platforms, they currently lack privacy tooling. Third, automated compliance tools could bridge legal complexity and research practice. Provenance tracking systems that generate GDPR-compliant documentation while qualifying for regulatory safe harbours \citep{longpre2024position} would transform PETLP's decision trees into executable compliance pipelines. A forthcoming \texttt{RedditHarbor} toolkit demonstrates one path forward -- automating PETLP's compliance logic specifically for Reddit research. These developments recognise that responsible AI research requires not just frameworks but ecosystems where institutional support, technical infrastructure, and automated tools converge to make compliance achievable rather than aspirational.

\section{Conclusion}

PETLP repositions privacy and compliance from constraints to design principles that enhance research rigour and public trust. Through detailed legal analysis and practical implementation, we demonstrate that responsible research practice strengthens rather than compromises scientific innovation -- creating conditions for sustainable, trustworthy AI development that withstands regulatory scrutiny and maintains social licence. We position PETLP not as a definitive solution but as a structured foundation for ongoing dialogue about how AI research can serve societal benefit whilst respecting individual privacy in an era where these values increasingly conflict.

\section*{Acknowledgments}
Dr. Giorgos D. Vrakas was supported by a postdoctoral fellowship within the ONISILOS MSCA COFUND program at the University of Cyprus, funded by the European Union’s Horizon 2020 research and innovation programme under the Marie Skłodowska-Curie Grant Agreement No. 101034403.

\section*{Legal Disclaimer} This paper is intended for informational and academic purposes only. The statements made regarding the law, ethics, and logistics of social media data practices are based on general understanding and research as of the publication date. This document does not constitute legal advice, nor does it create an attorney-client relationship. Laws and regulations surrounding data extraction, processing, storage, and presentation may vary by jurisdiction and are subject to change. The authors make no representations or warranties regarding the accuracy, completeness, or applicability of the information contained herein. Readers are encouraged to consult qualified Data Protection Officers or legal professionals to obtain advice specific to their individual circumstances before implementing any of the approaches described. The authors disclaim all liability for any actions taken based on the content of this paper. 

\bibliography{aaai25}

\newpage
\appendix

\section{GDPR Principles for Social Media Research}

\paragraph{Social Media Data as Personal Data}

The GDPR defines personal data broadly as `any information relating to an identified or identifiable natural person' (Article 4). This technology-neutral approach intentionally encompasses diverse forms of information across various mediums, ensuring the regulation remains applicable regardless of technological developments. Central to this definition are the concepts of \textit{identification} and \textit{identifiability}. A data subject is considered `identified' when they can be directly or indirectly singled out from others based on available information, even if their name is not explicitly known \cite{WP29Opinion2007}. Likewise, a person is `identifiable' when identification becomes possible through the combination of disparate pieces of information, even if this possibility is not immediately apparent \cite{WP29Opinion2007}. As technological capabilities advance, particularly in the field of AI, the threshold for identifiability continues to expand, broadening what qualifies as personal data in research contexts.

Social media data, therefore, fundamentally qualifies as personal data under the GDPR, irrespective of whether it is publicly accessible. \textit{The determining factor lies not in the data’s availability but in its potential to identify individuals directly or indirectly}. This interpretation is reinforced by both the Article 29 Data Protection Working Party (now the European Data Protection Board, or EDPB), which explicitly highlighted the requirement for DPIAs when processing social media data \cite{WP29DPIAGuidelines}, and by the Court of Justice of the European Union in \textit{Maximilian Schrems v. Meta Platforms Ireland (C-446/21)} \citeyear{Schrems2024}.

This consideration is particularly significant when recognising that social media data frequently contains or enables the inference of `special categories of personal data' under Article 9 GDPR. Such data includes information revealing or permitting the deduction of sensitive characteristics such as health status, racial or ethnic origin, religious beliefs, or political opinions. Recent CJEU rulings \cite{OT2022, Meta2023, Lindenapotheke2023} have clarified that explicit disclosure of such characteristics is not required to trigger Article 9 protections. \textit{Instead, the mere possibility of inferring or deducing protected attributes -- regardless of the accuracy of such inferences -- activates these heightened legal safeguards}. This principle carries profound implications for applied AI research, where advanced models \textit{may} infer sensitive characteristics from seemingly innocuous patterns of language use, user interactions, or posting behaviour. AI profiling techniques can similarly transform ordinary personal data into special category data when combined with other information or contextual analysis \cite{WP29ProfilingGuidelines}.

These risks are amplified by the capabilities of modern AI systems, particularly LLMs, which may inadvertently memorise or reproduce personal data from their training sets, including special category information \cite{EDPB2024Opinion28}. Such unintended inferences can occur even when researchers do not explicitly design models to process sensitive attributes. This demands that researchers account not only for their intended analytical outcomes but also for the \textit{latent capacities of their models to generate unforeseen inferences about individuals}.

While the GDPR provides a pathway for processing special category data under Article 9(2)(j) for scientific research purposes, this legal permission comes with the expectation of proportionate and robust safeguards. Responsible AI development, therefore, requires the implementation of privacy-preserving techniques, the execution of thorough risk assessments, and the establishment of governance protocols that protect data subjects' rights by design and by default.

\paragraph{The Researcher as Joint Controller}

In the context of AI research using social media data, understanding who bears responsibility for data protection is crucial. The GDPR defines a controller as `the natural or legal person, public authority, agency or other body which, alone or jointly with others, determines the purposes and means of the processing of personal data' (Article 4(7)). For AI researchers, this means identifying who decides the \textit{why} and \textit{how} of data processing decisions that fundamentally shape the research.

Formally, the research institution typically acts as the controller \cite[p. 3]{EDPBControllerProcessor2020}. However, controllership is a functional concept determined by actual influence over data processing decisions, not merely formal designation \cite{EDPBControllerProcessor2020}. In practice, AI researchers decides data collection methodologies, transformation techniques, and model training approaches that directly impact data subjects' privacy. These decisions -- such as which data fields to collect, what anonymisation techniques to employ, or how to balance model performance against privacy preservation -- are integral to determining the \textit{means} of processing.

This paper therefore advocates viewing researchers as joint controllers with their institutions. Joint controllership emerges when different parties jointly determine the purposes and means of processing through complementary, necessary decisions \cite{EDPBControllerProcessor2020}. By conceptualising AI researchers as joint controllers, we establish a framework where privacy considerations permeate the entire pipeline -- from initial data extraction to model deployment -- rather than remaining abstract institutional obligations. 

\paragraph{Data Protection by Design and Impact Assessments}

Article 24 of the GDPR (`Responsibility of the Controller') operationalises the accountability principle established in Article 5(2), requiring controllers to implement appropriate technical and organisational measures to ensure and demonstrate compliance. This accountability framework adopts a risk-based approach, with the DPIA serving as its primary mechanism when processing is `likely to result in a high risk to the rights and freedoms of natural persons' (Article 35).

For AI researchers working with social media data, DPIAs are effectively mandatory. The former Article 29 Working Party identified nine criteria signalling high-risk processing \cite{WP29DPIAGuidelines, WP29ProfilingGuidelines}, and AI research involving social media data routinely triggers multiple indicators: evaluation or scoring of individuals (sentiment analysis, user classification), processing of sensitive data (health discussions, political opinions), large-scale processing (training LLMs), dataset combination (linking profiles across platforms), and deployment of innovative technologies (deep learning, generative models).

A compliant DPIA must satisfy specific requirements under Article 35(7): (1) Provide a systematic description of the processing (35(7)(a)), documenting the nature, scope, context, and purposes -- including data categories, recipients, retention periods, and technical infrastructure; (2) Assess necessity and proportionality (35(7)(b)), demonstrating how processing achieves legitimate purposes whilst implementing Article 5 data protection principles; and (3) Identify and manage risks (35(7)(c)), evaluating threats from illegitimate access, undesired modification, or data loss, whilst determining appropriate mitigations. Throughout this process, researchers must seek DPO advice (Article 35(2)) and consider consulting data subjects where appropriate (Article 35(9)).

Importantly, DPIAs are not static or one-off exercises. They are \textit{living} documents that must be revisited and updated when there are changes to the processing activities, models, or data sources \cite[p. 9]{WP29DPIAGuidelines} -- an aspect particularly relevant in iterative AI development cycles where models are frequently re-trained or re-purposed. While the GDPR does not mandate full public disclosure of DPIAs, publishing a summary of the assessment and its key findings can serve as a powerful mechanism for fostering public trust and demonstrating organisational transparency.

\paragraph{Establishing a Legal Basis for Processing}

Before any processing of personal data can commence, researchers must identify and document a valid legal basis under Article 6 GDPR -- a requirement that applies regardless of whether data is publicly accessible or seemingly non-sensitive. For social media research, two legal bases prove particularly relevant: public interest (Article 6(1)(e)) and legitimate interests (Article 6(1)(f)). 

The availability of these bases depends fundamentally on the organisational context. Article 6(1)(e) permits processing when necessary for the performance of a task carried out in the public interest \textit{or} in the exercise of official authority vested in the controller. Both public authorities and private entities can rely on this basis when performing legally-established public interest tasks. Universities, as public authorities whose constitutional documents typically define research as part of their role, generally rely on public interest for research activities. For example, University College London adopted a Statement of Tasks in the Public Interest \cite{ucl2018publictask} defining its mission as `facilitating and carrying out research in any field'. This broad mandate encompasses even research entirely funded by private companies, with UCL stating that most research activities fall under the `public task' condition. Notably, Article 6(1)(f) explicitly states that legitimate interests `shall not apply to processing carried out by public authorities in the performance of their tasks'. This prohibition means that whilst private companies and independent research organisations may invoke legitimate interests, public authorities conducting research within their official remit cannot. Researchers must distinguish clearly between their broader \textit{interest} (the overarching objective, such as improving mental health detection) and their specific \textit{purpose} (the concrete processing activity, such as analysing Reddit posts to train a classification model).

Each legal basis imposes distinct procedural requirements. Public interest processing under Article 6(1)(e) requires demonstrating a clear basis in law for the task or authority, which need not be a specific statutory power but must be grounded in constitutional documents, charters, or legal frameworks. The processing must be \textit{necessary} -- meaning no less intrusive means exist to achieve the objective. When processing special category data, researchers must additionally satisfy Article 9(2)(j) by implementing Article 89(1) safeguards, including data minimisation, pseudonymisation where feasible, and ethical oversight. In contrast, legitimate interests under Article 6(1)(f) -- available to private entities and public authorities acting outside their official tasks -- requires a documented \textit{Legitimate Interest Assessment} (LIA) that evaluates: (1) the legitimacy of the interest pursued; (2) the necessity of processing, confirming no less intrusive means exist; and (3) a balancing test weighing the controller's interests against data subjects' rights and freedoms. Commercial research organisations, private consultancies, and industry research divisions typically rely on this basis, as they cannot invoke public task provisions. As emphasised in EDPB Guidelines 1/2024, only if this balancing test favours the controller may processing proceed.

Regardless of the chosen legal basis, the GDPR affords scientific research privileged status through multiple flexibilities designed to facilitate research whilst maintaining privacy protections. Recital 159 mandates a broad interpretation of `scientific research', encompassing `technological development and demonstration, fundamental research, applied research and privately funded research', provided it follows recognised methodological and ethical standards \cite{edpb2021secondaryuse}. These privileges include purpose compatibility for secondary use (Article 5(1)(b)), extended retention periods (Article 5(1)(e)), transparency exemptions when individual notification requires disproportionate effort (Article 14(5)(b)), and conditional limitations on erasure and objection rights where these would seriously impair research objectives (Articles 17(3)(d) and 21(6)). However, these flexibilities do not exempt researchers from implementing appropriate Article 89(1) safeguards -- rather, they represent a calibrated balance between enabling research and protecting individual privacy through technical and organisational measures.

\newpage
\section{WP29 Criteria for High-Risk Processing Requiring DPIA}
\label{app:WP29DPIAcriteria}

The GDPR does not provide a precise definition of `high risk' processing that mandates a DPIA. Instead, Article 35(3) offers non-exhaustive examples of processing operations likely to result in high risk:

\begin{itemize}
\item Systematic and extensive evaluation of personal aspects through automated processing, including profiling, which produces legal or similarly significant effects (Recital 71 clarifies this includes analysing or predicting aspects concerning performance at work, economic situation, health, personal preferences or interests, reliability or behaviour, location or movements);
\item Large-scale processing of special category data under Article 9(1) or criminal conviction data under Article 10 (Recital 75 specifies data which reveal racial or ethnic origin, political opinions, religion or philosophical beliefs, trade union membership, and the processing of genetic data, data concerning health or data concerning sex life);
\item Systematic monitoring of publicly accessible areas on a large scale.
\end{itemize}

To provide more concrete guidance, the Article 29 Data Protection Working Party (now the European Data Protection Board) identified nine criteria in its Guidelines on DPIA \cite{WP29DPIAGuidelines} to help controllers determine when processing is `likely to result in a high risk':

\begin{enumerate}
\item \textbf{Evaluation or scoring}: Processing that involves assessing or scoring individuals based on aspects such as behaviour, interests, or health, particularly through profiling. This can significantly affect individuals' rights, freedoms, or opportunities.

\item \textbf{Automated decision-making with legal or similar significant effect}: Processing that results in automated decisions producing legal effects or similarly significantly affecting individuals, such as automatic refusal of credit applications or e-recruiting without human intervention.

\item \textbf{Systematic monitoring}: Ongoing or systematic observation of individuals, particularly in publicly accessible areas where individuals may be unaware of the data collection or unable to avoid it.

\item \textbf{Sensitive data or data of a highly personal nature}: Processing special categories of data as defined in Article 9, as well as other highly personal data (such as electronic communications, location data, or financial data) where misuse could seriously impact individuals' rights and freedoms.

\item \textbf{Data processed on a large scale}: Operations involving vast numbers of data subjects, high volumes of data, extended duration, or wide geographic scope. Whilst the GDPR does not define ``large scale'', Recital 91 suggests considering the number of data subjects, volume of data, duration, and geographical extent.

\item \textbf{Matching or combining datasets}: Merging data from different sources or collected for different purposes, particularly in ways that exceed individuals' reasonable expectations of how their data would be used.

\item \textbf{Data concerning vulnerable data subjects}: Processing data of individuals who may be unable to freely consent or object due to an imbalance of power, including children, employees, elderly individuals, patients, or those with mental health conditions.

\item \textbf{Innovative use or applying new technological or organisational solutions}: Deployment of novel technologies (such as AI systems, IoT devices, or biometric technologies) or innovative applications of existing technologies that may introduce unknown or heightened privacy risks.

\item \textbf{Processing that prevents data subjects from exercising a right or using a service or contract}: Operations that directly limit individuals' ability to access services, exercise rights, or enter contracts, such as credit scoring used to deny loans or blacklisting databases.
\end{enumerate}

\subsection{Application to Social Media Research}

The WP29 guidelines establish that processing meeting \textit{two or more criteria} typically requires a DPIA. Moreover, the more criteria are met by the processing, the more likely it is to present a high risk to the rights and freedoms of data subjects, and therefore to require a DPIA \cite[p. 11]{WP29DPIAGuidelines}. This assessment operates independently of any measures which the controller envisages to adopt.

Notably, in the context of social media data, the WP29 explicitly identified two scenarios requiring DPIAs:
\begin{itemize}
\item Gathering of public social media data for generating profiles
\item Storage for archiving purposes of pseudonymised personal sensitive data concerning vulnerable data subjects of research projects or clinical trials
\end{itemize}

Given that social media AI research routinely involves multiple criteria -- including profiling (criterion 1), sensitive data inference (criterion 4), large-scale processing (criterion 5), dataset combination (criterion 6), potential inclusion of vulnerable subjects (criterion 7), and innovative AI technologies (criterion 8) -- DPIAs are effectively mandatory for most social media research projects. 

Even where uncertainty exists, the WP29 recommends that a DPIA is
carried out nonetheless as it is a useful tool to help controllers comply with data protection law \cite[p. 8]{WP29DPIAGuidelines}.

\newpage
\section{Integrating the EU Artificial Intelligence Act into PETLP}
\label{app:EUAIAct}

Although the most visible transparency duties of the EU Artificial Intelligence Act (AI Act) -- such as content-origin disclosure and watermarking for generative systems (Article 50) -- arise when a model or dataset is \emph{shared} with external audiences, the Regulation embeds legally enforceable controls across \emph{all} stages of the PETLP pipeline. This appendix examines how the AI Act's requirements map onto our framework and addresses a critical tension between differential privacy techniques and data quality obligations.

\subsection{Mapping AI Act Requirements to PETLP Stages}

\paragraph{Extract: Provenance, Bias, and Lawful Sourcing}
Any AI system that will ultimately be \emph{placed on the market} or \emph{put into service} within the Union must, from its inception, operate under a risk-management framework (Article 9) and data-governance controls (Article 10). For providers who scrape or ingest Reddit, X, or other platform corpora, this means documenting collection processes, representativeness, potential biases and mitigation strategies before the first byte is written to disk. If the resulting model is likely to be classed as \emph{general-purpose AI} (GPAI) (following Article 51), the provider must additionally prepare a `sufficiently detailed' summary of the training data categories (Annex XI, Article 53).

\paragraph{Transform: Quality Assurance and Design-time Safeguards}
Cleaning, labelling, augmenting, or otherwise modifying social-media corpora must be logged and justified as part of the same risk-management system; Article 10(3) explicitly requires that training, validation and test sets remain \emph{representative, error-free, and complete as far as possible}. High-risk developers must also compile technical documentation (Article 11) and human-oversight protocols (Article 14) durning development. These duties dovetail with the default-DPIA workflow already mandated by the GDPR in our privacy-by-design approach.

\paragraph{Load: Secure Logging and Cross-border Collaboration}
During the Load phase, researchers configure storage, access controls and intra-consortium sharing. Article 12 obliges providers of high-risk systems to maintain \emph{automatic logs} `to ensure a level of traceability' commensurate with the risk. When data travel outside the EEA, GDPR transfer rules (Articles 44--49) apply in parallel, and the AI Act's ex-territorial reach ensures that non-EU collaborators who are `providers' or `deployers' must also respect these logging and documentation requirements.

\paragraph{Present: User-facing Transparency and Output Liability}
Only at the dissemination stage do the Act's familiar consumer-visible obligations take centre stage: disclosure to users that they are interacting with an AI system (Article 13), watermarking or labelling of AI-generated audio-visual content (Article 52), and, for certain use-cases, a Fundamental Rights Impact Assessment by the deployer (Article 27). In addition, GPAI providers must publish the Annex XI data-summary and co-operate with the newly created AI Office (Article 53).

\paragraph{Research Exemption and Practical Nuance}
Article 2(6) excludes AI systems `developed and put into service for the sole purpose of scientific research and development' from the Act's operational scope. \emph{Academic} teams who never release their models, weights or APIs therefore remain governed solely by the GDPR-IP-contract bundle analysed earlier. \emph{Industrial} R\&D units, however, must re-enter the Act's compliance track the moment a prototype is commercialised or otherwise `put into service'. Likewise, an open-source release escapes the GPAI chapter only if the model poses no `systemic risk' (Article 53(2)).

\paragraph{Implication for PETLP}
The AI Act thus acts as a second, vertically layered compliance spine: its Article 9--12 duties map onto Extract, Transform and Load, whilst its Article 13 \& 52 duties align with Present. Integrating these checkpoints into the \emph{default-DPIA} that guides PETLP ensures that social-media researchers -- academic or commercial -- address EU regulatory demands \emph{before} dissemination, rather than treating them as an afterthought.

\subsection{Reconciling Differential Privacy with Article 10(3) Data-Quality Duties}
\label{subsec:dp_vs_a10}

Article 10(3) AI Act requires that training, validation and test datasets be `relevant, sufficiently representative and, to the best extent possible, free of errors and complete' in view of the system's intended purpose. At first sight, the calibrated noise that constitutes \emph{differential privacy} appears to inject \emph{errors} into the data, thereby frustrating compliance. A closer reading shows the relationship to be one of \emph{managed trade-off} rather than outright conflict:

\begin{enumerate}
\item \textbf{Risk-based standard:} `Free of errors' \ldots \emph{to the best extent possible} is a reasonableness clause, not a strict liability rule. Recital 67 that underpins Article 10(3) recognises that ensuring representativeness or completeness may require \emph{balancing} competing constraints, including privacy protection.

\item \textbf{Scope of the duty:} Article 10(3) speaks to the \emph{statistical properties} of the dataset \emph{as a whole}. DP noise is, by design, zero-mean and unbiased; it degrades individual-level fidelity whilst preserving aggregate distributions within a quantifiable privacy budget. Hence a DP-protected corpus can remain `sufficiently representative' if the privacy parameter~$\varepsilon$ is chosen so that downstream model performance meets documented acceptance criteria. The Act's risk-management system (Article 9) provides the forum for justifying that calibration.

\item \textbf{Complementarity with GDPR minimisation:} Differential privacy operationalises GDPR \emph{data minimisation} by guaranteeing that no single record has material influence on the output; the AI Act does not override this principle but obliges providers to \emph{evidence} that the resulting utility remains adequate.

\item \textbf{Documentation and testing:} The provider must (i) log the DP mechanism and chosen~$\varepsilon$ in the Article 11 technical documentation; (ii) demonstrate through validation metrics that the privacy noise does not induce prohibited bias or unacceptable accuracy loss (Article 9); and (iii) record this analysis in the default-DPIA so that supervisory authorities can audit the privacy-utility trade-off.
\end{enumerate}

In summary, Article 10(3) does not prohibit differential privacy; it requires providers to show, through the Act's risk-management and documentation pillars, that the chosen level of noise still yields a dataset \emph{fit for purpose}. Properly calibrated DP therefore serves both GDPR minimisation and AI-Act data-quality goals, provided the trade-off is transparently justified and empirically validated. This reconciliation exemplifies the broader principle underlying PETLP: that privacy-preserving techniques and regulatory compliance can be mutually reinforcing when thoughtfully integrated into research design.

\subsection{Additional Resources}

In July 2025, the Commission introduced three instruments to support GPAI model development and deployment:

\begin{itemize}
\item \textbf{Guidelines on the scope of obligations for providers of GPAI models} \cite{EU_Guidelines_GPAI2025}: Clarifies which actors along the AI value chain must comply with GPAI obligations, helping researchers and developers determine their regulatory status.

\item \textbf{GPAI Code of Practice} \cite{EU_Code_GPAI2025}: A voluntary compliance tool developed by independent experts, offering practical guidance on meeting transparency, copyright, and safety obligations under the AI Act.

\item \textbf{Template for public summary of training content} \cite{EU_AI_Template2025}: Standardises disclosure of training data sources, including large datasets and primary domains, whilst documenting data processing aspects to facilitate rights exercise under EU law.
\end{itemize}

These instruments complement the regulatory framework, particularly valuable for researchers navigating the intersection of AI Act requirements with the PETLP framework. While not legally binding (except the template requirement under Article 53), they represent authoritative interpretations that supervisory authorities are likely to reference when assessing compliance.

\newpage
\section{Decision Trees for Legal Compliance in Social Media AI Research}

This appendix provides nine decision trees operationalising the legal requirements discussed throughout this paper. Each addresses a specific compliance challenge in social media research. 

\textit{Note: Trees contain detailed pathways optimised for digital viewing. Zoom in to see all decision nodes, citations, and guidance clearly.}

\textbf{Figure 2: Determining Controller Relationships} -- Identifies controller-processor relationships in collaborative research, distinguishing joint controllers, independent controllers, and processors. Essential for allocating GDPR responsibilities under Articles 26 and 28.

\textbf{Figure 3: GDPR Legal Basis Selection} -- Systematic approach to selecting legal bases, guiding through consent, public task, and legitimate interest assessments. Differentiates pathways by institutional status and research purpose.

\textbf{Figure 4: Research Organisation Qualification} -- Determines eligibility for DSM Article 3's mandatory text and data mining exception through assessment of institutional status, research purpose, and commercial influence. Critical for understanding whether platform terms can override statutory research rights.

\textbf{Figure 5: Platform Restrictions versus Legal Rights} -- Clarifies when terms of service yield to statutory research exceptions, particularly for DSM Article 3 organisations. Addresses practical enforcement considerations.


\textbf{Figure 6: Data Extraction Compliance} -- Integrates pre-determined researcher status with extraction method selection, then maps GDPR notification requirements based on data accessibility. Addresses special category data considerations and Article 14(5)(b) disproportionate effort assessments.

\textbf{Figure 7: Transformation Stage Safeguards} -- Navigates the dual challenges of copyright compliance (where each transformation creates reproductions) and privacy engineering (implementing minimisation and anonymisation). Highlights the extreme difficulty of achieving true anonymisation with social media data.

\textbf{Figure 8: Storage and Retention Compliance} -- Maps storage decisions to security requirements and international transfers under Articles 44-49 GDPR. Retention policies differ by DSM Article 3 qualification.

\textbf{Figure 9: Model Training Risk Assessment} -- Focuses on testing AI models for data leakage before deployment. Emphasises heightened obligations for public model releases versus internal deployments.

\textbf{Figure 10: Research Output Distribution} -- Guides compliant dissemination of research outputs from papers to datasets, addressing tensions between open science mandates and privacy/copyright restrictions. Differentiates pathways based on researcher status and output type, incorporating platform-specific limitations.

\begin{figure*}[!htbp]
\centering
\includegraphics[height=\textheight]{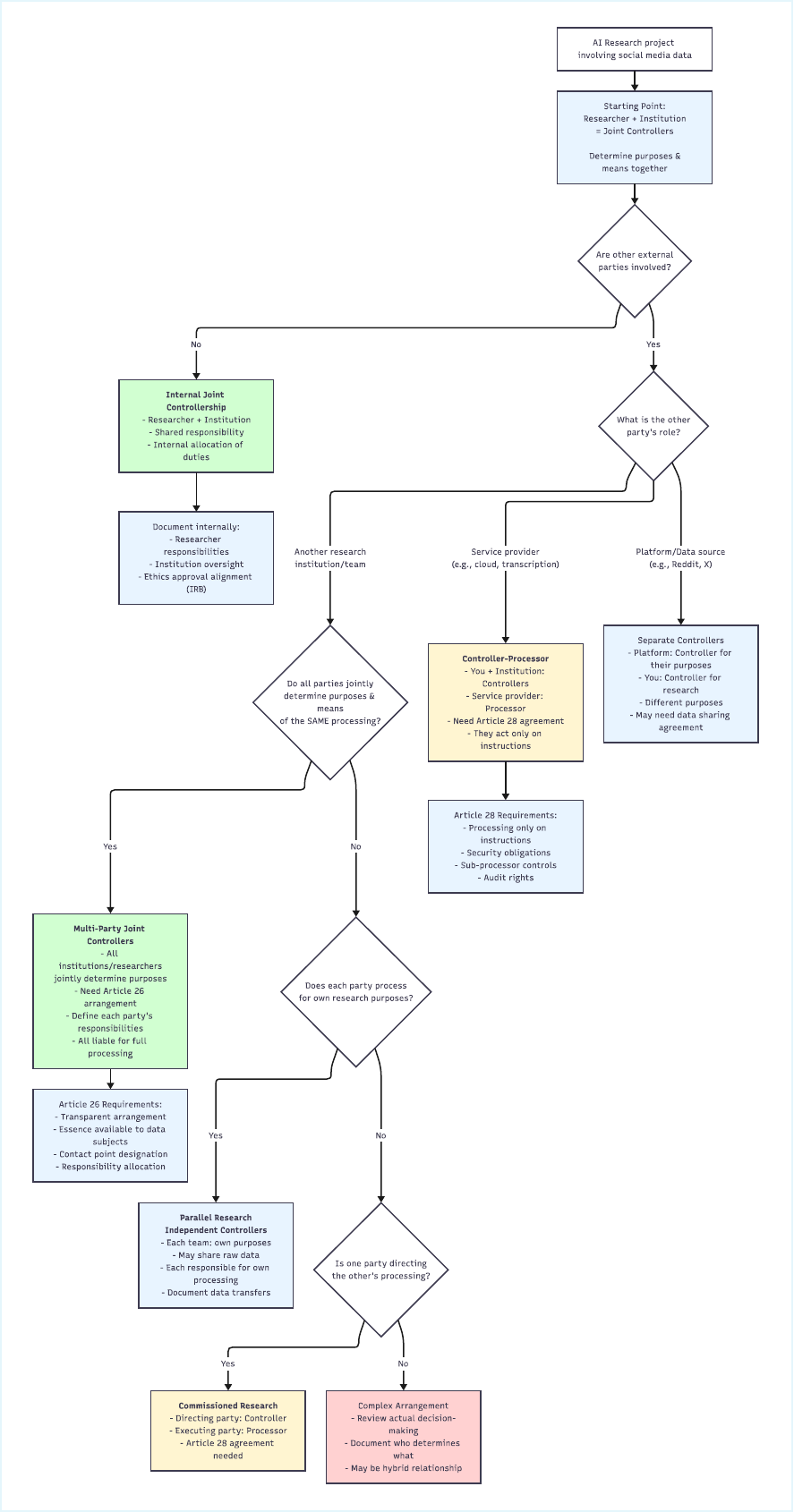}
\caption{\textbf{Determining Controller Relationships in AI Research Projects}. \textit{Identifying joint controllers, processors, and independent controllers for GDPR compliance}}
\end{figure*}

\begin{figure*}[!htbp]
\centering
\includegraphics[height=\textheight]{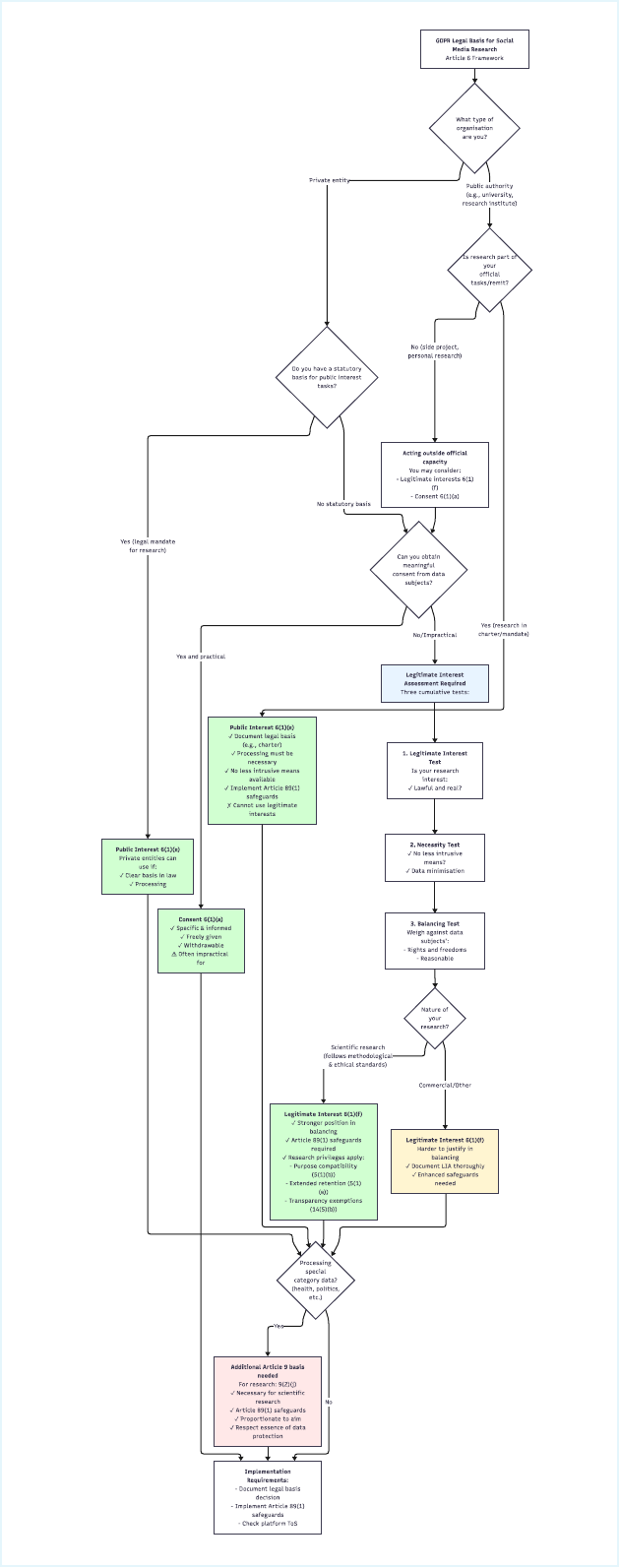}
\caption{\textbf{GDPR Legal Basis Selection for Social Media Research}. \textit{Navigating consent, legitimate interest, and public task for AI research projects}}
\end{figure*}

\begin{figure*}[!htbp]
\centering
\includegraphics[height=\textheight]{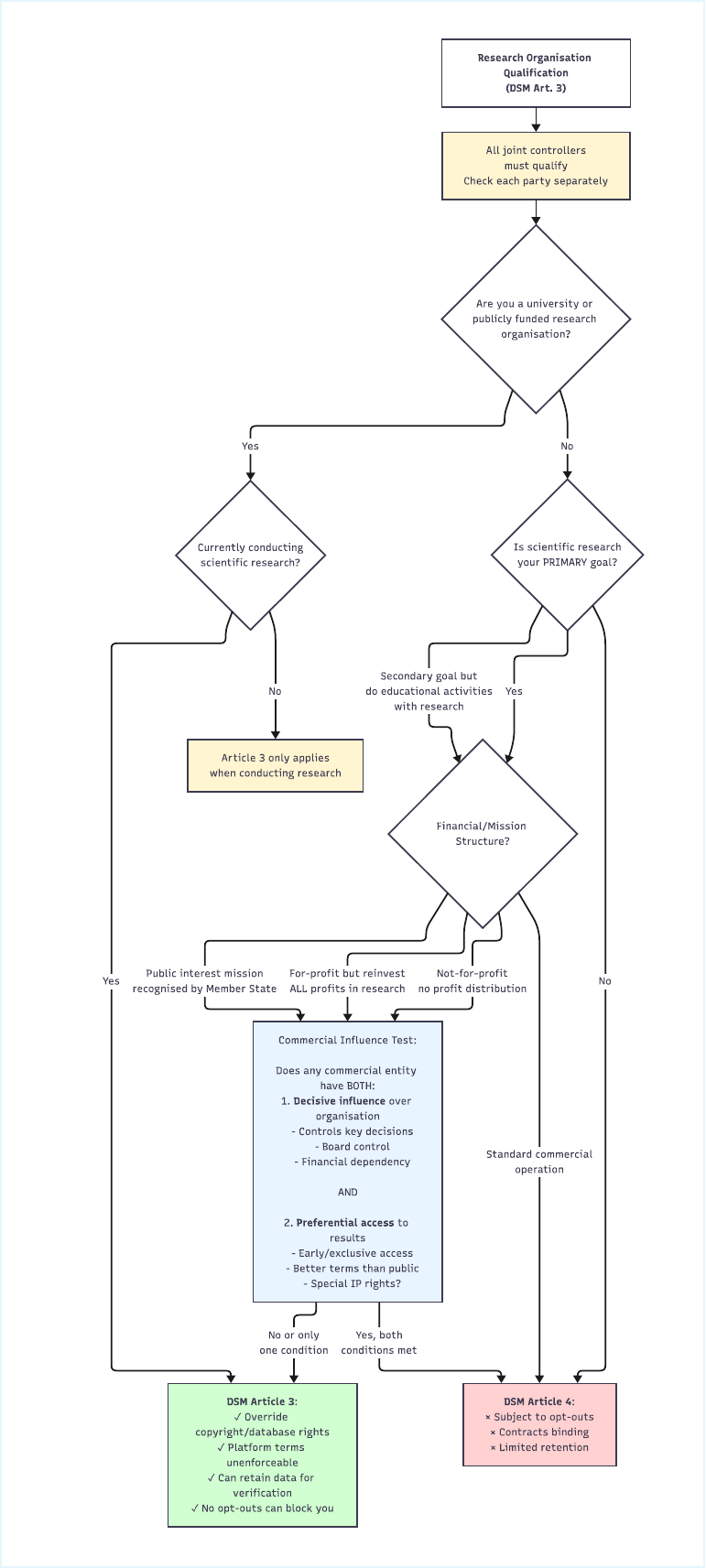}
\caption{\textbf{Research Organisation Qualification Under DSM Article 3}. \textit{Determining eligibility for enhanced text and data mining rights}}
\end{figure*}

\begin{figure*}[!htbp]
\centering
\includegraphics[height=\textheight]{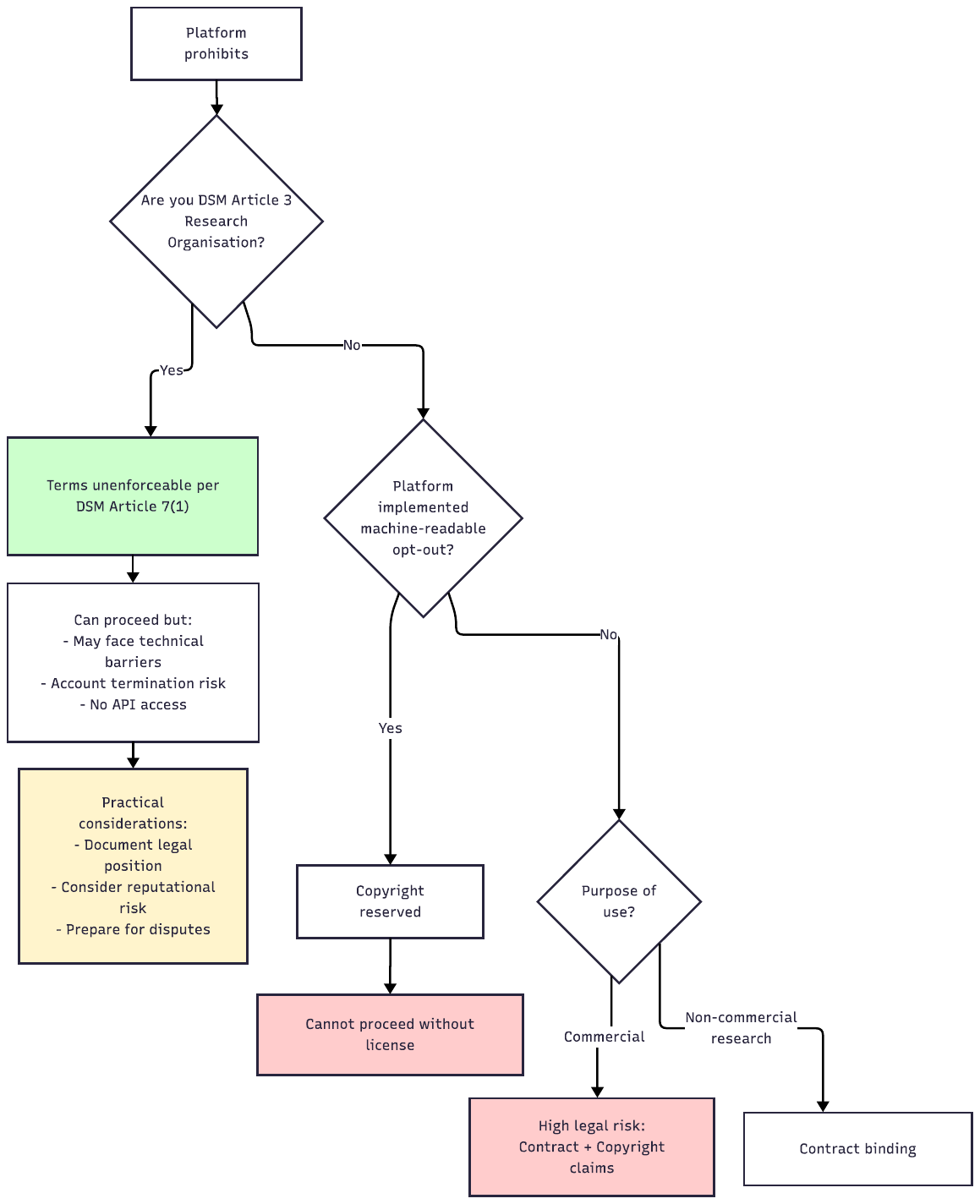}
\caption{\textbf{Platform Restrictions versus Legal Rights in AI Research}. \textit{When terms of service conflict with statutory research exemptions}}
\end{figure*}


\begin{figure*}[!htbp]
\centering
\includegraphics[height=\textheight]{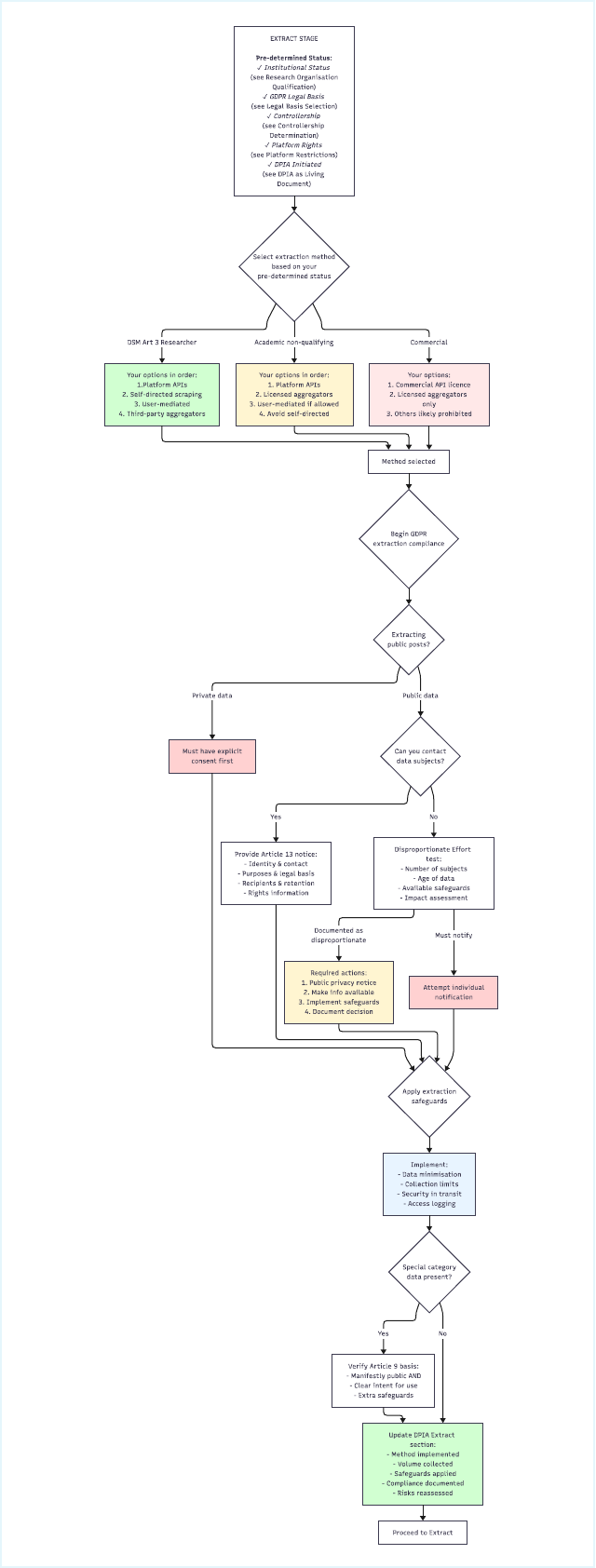}
\caption{\textbf{Data Extraction Compliance Framework}. \textit{Method selection and GDPR notification requirements for social media data collection}}
\end{figure*}

\begin{figure*}[!htbp]
\centering
\includegraphics[height=\textheight]{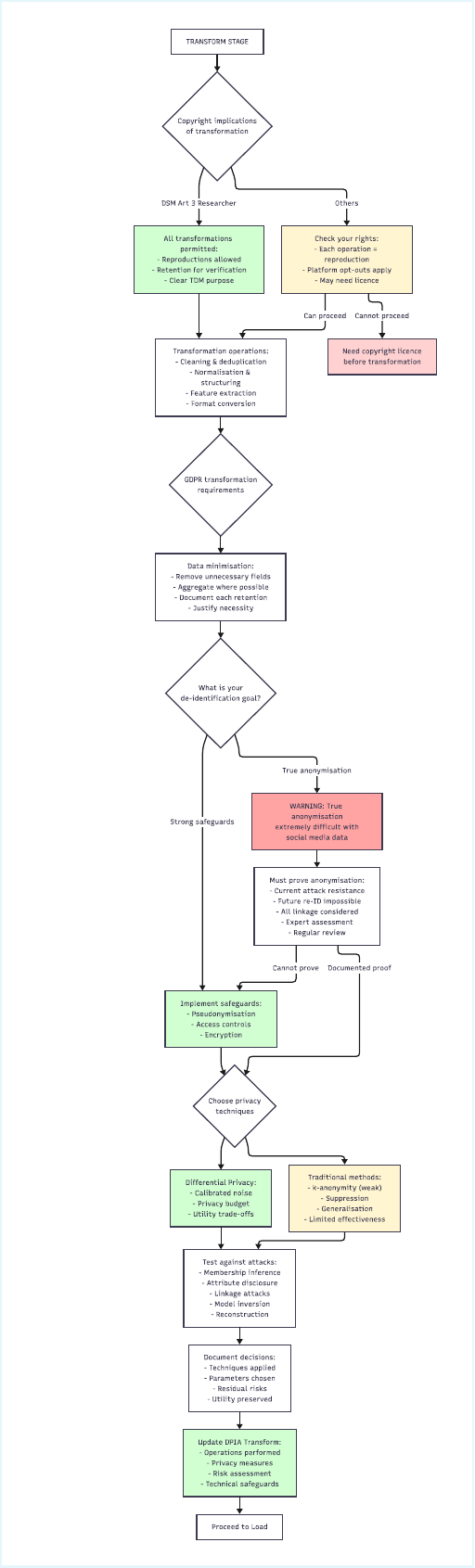}
\caption{\textbf{Transformation Stage Privacy Safeguards}. \textit{Implementing de-identification techniques and assessing anonymisation claims}}
\end{figure*}

\begin{figure*}[!htbp]
\centering
\includegraphics[height=\textheight]{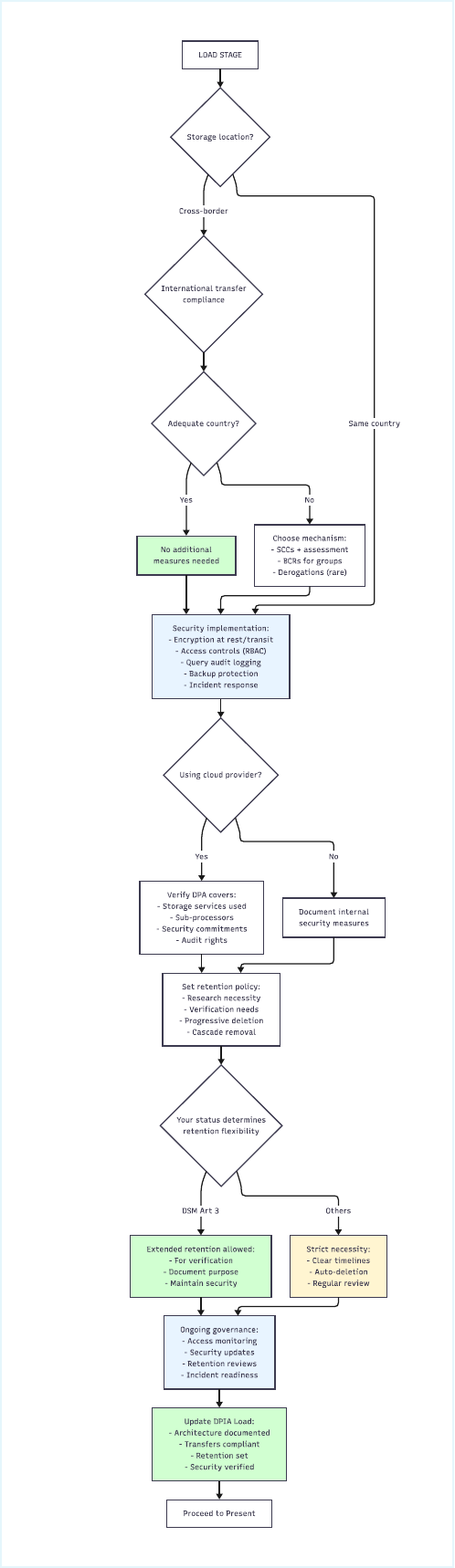}
\caption{\textbf{Data Storage and Retention Compliance}. \textit{Cross-border transfers, security measures, and retention policies}}
\end{figure*}

\begin{figure*}[!htbp]
\centering
\includegraphics[height=\textheight]{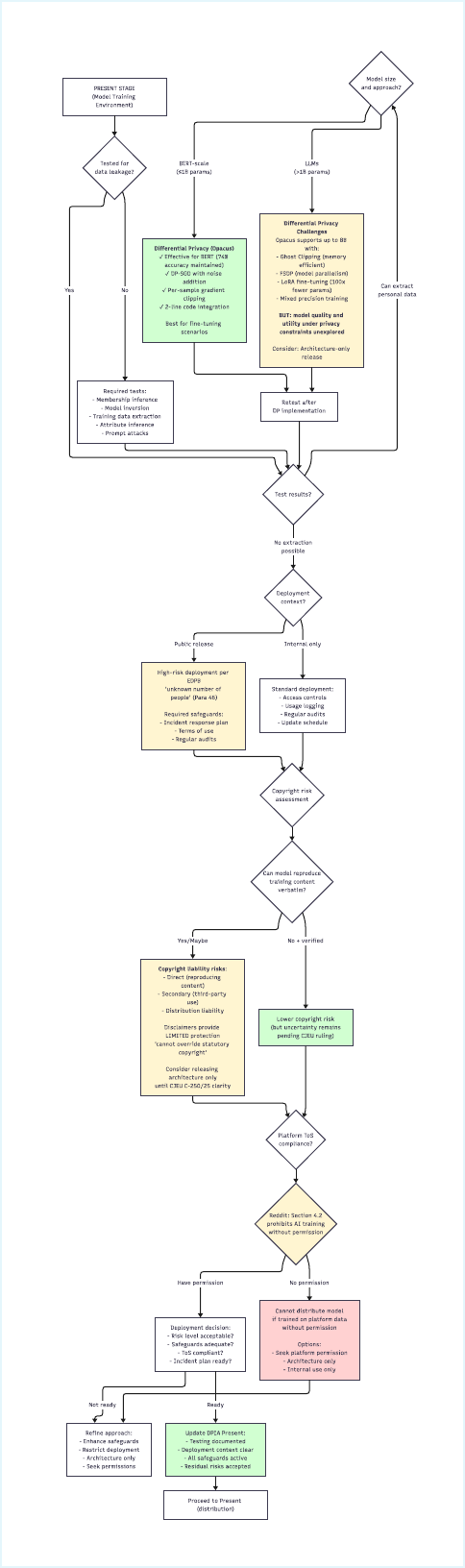}
\caption{\textbf{Model Training and Deployment Risk Assessment}. \textit{Testing for data leakage and implementing deployment safeguards}}
\end{figure*}

\begin{figure*}[!htbp]
\centering
\includegraphics[height=\textheight]{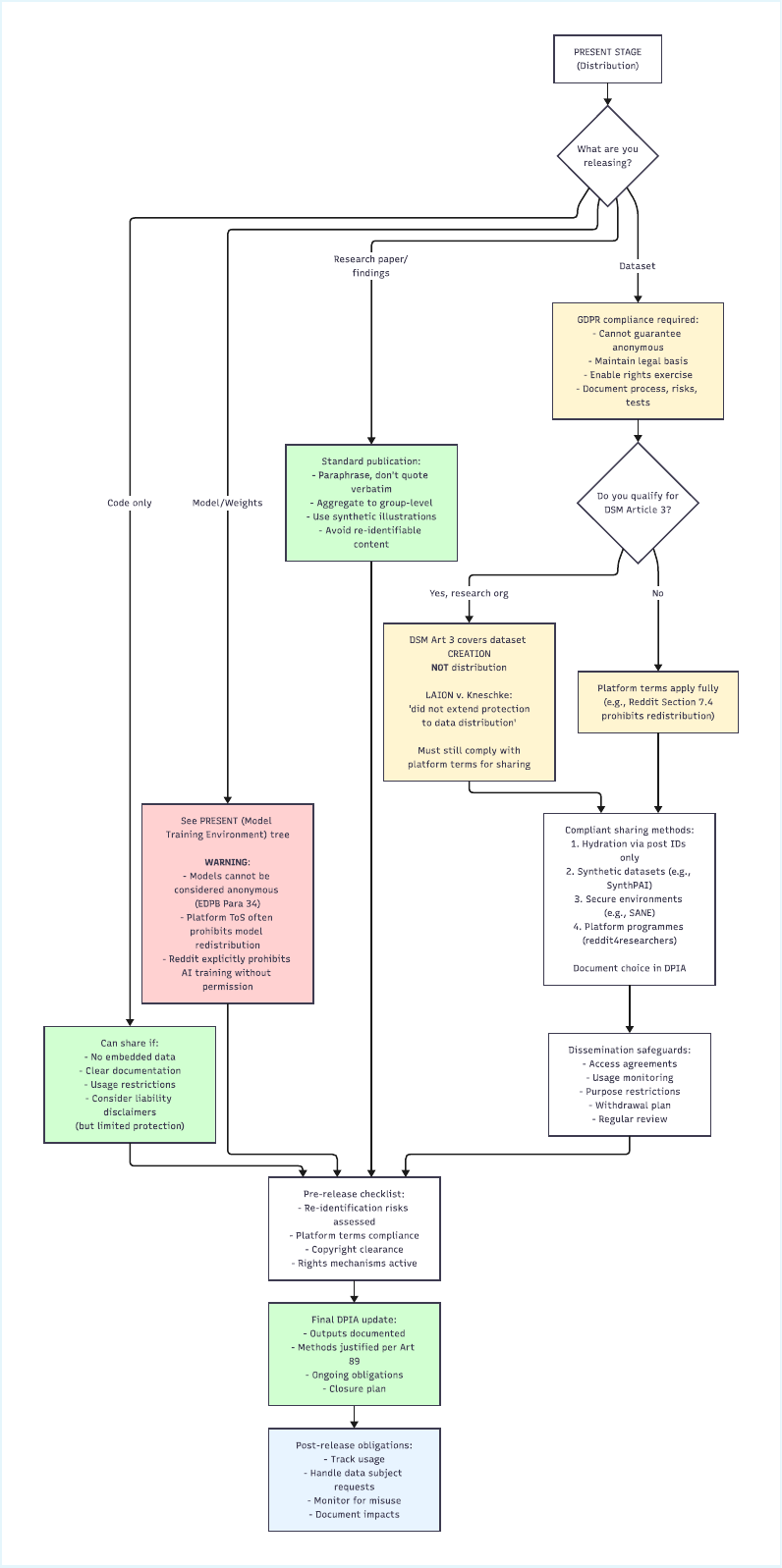}
\caption{\textbf{Research Output Distribution Decision Framework}. \textit{Sharing papers, datasets, models, and code under the GDPR and copyright law}}
\end{figure*}

\clearpage
\section{Comprehensive Risk Assessment Example}

This section provides a practical walk-through of the PETLP framework applied to political discourse analysis.

\subsection{Research Context}

\noindent\textbf{Researcher:} Doctoral student at a European university, Department of Political Science

\noindent\textbf{Research Question:} How do political discussion patterns differ across ideological communities on Reddit?

\noindent\textbf{Methodology:} Fine-tune \texttt{Llama-3.2-3B} to classify political stance and analyse discourse patterns across 100,000 Reddit posts from diverse political subreddits.

\noindent\textbf{Intended Outputs:} Academic paper, fine-tuned model for research community, aggregated statistics on political discourse patterns.

\subsection{Pre-Registration and DPIA Initiation}

\subsubsection{Pre-Registration Components}

\begin{enumerate}
\item \textbf{Research Hypotheses:}
\begin{itemize}
\item H1: Political communities exhibit distinct linguistic patterns correlating with ideological positioning
\item H2: Cross-ideological discussions show increased linguistic complexity compared to within-group discussions
\end{itemize}

\item \textbf{Study Design:}
\begin{itemize}
\item Observational study of naturally occurring Reddit discussions
\item Comparative analysis across three political subreddits: r/politics, r/Conservative, r/democrats
\item Time period: January 2024 -- December 2024
\end{itemize}

\item \textbf{Data Collection Plan:}
\begin{itemize}
\item Target: 100,000 posts (balanced across subreddits)
\item Method: Reddit API (free tier, 100 queries per minute)
\item Fields: selftext (post content), created\_utc, subreddit, score (no usernames)
\end{itemize}

\item \textbf{Model Training Design:}
\begin{itemize}
\item Base model: \texttt{Llama-3.2-3B} (chosen for efficiency and academic accessibility)
\item Fine-tuning approach: Political stance classification (3-class: left, centre, right)
\item Training infrastructure: University GPU cluster
\end{itemize}

\item \textbf{Expected Outputs:}
\begin{itemize}
\item Quantitative analysis of linguistic patterns
\item Fine-tuned model for political stance detection
\item Methodological contribution on privacy-preserving political discourse analysis
\end{itemize}
\end{enumerate}

\subsubsection{Mapping Pre-Registration to Initial DPIA}

\begin{table}[h]
\centering
\small
\caption{Pre-Registration to DPIA Mapping}
\label{tab:dpia-mapping}
\begin{tabular}{p{2cm}|p{5.5cm}}
\hline
\textbf{Element} & \textbf{DPIA Assessment} \\
\hline
Hypotheses & Purpose: Scientific research; Legal basis: Art. 6(1)(e); Political opinions (Art. 9) \\
\hline
Study Design & 100k posts, 3 subreddits; Controller: University; Large-scale processing \\
\hline
Data Plan & No usernames, public only; Art. 14(5)(b) notification; API compliant \\
\hline
Model Design & Base model risks; Differential privacy; Secure infrastructure \\
\hline
Outputs & Medium re-identification risk; Model bias; Monitor usage \\
\hline
\end{tabular}
\end{table}

\subsection{Extract -- Acquiring Reddit Data Legally}

This phase has revealed how the LAION decision enables qualifying research organisations to invoke DSM Article 3's mandatory TDM exception, overriding platform prohibitions on scraping for scientific research. However, commercial researchers remain bound by platform terms and Article 4's opt-out provisions (like Reddit's robots.txt), whilst GDPR operates independently -- requiring all researchers to document legal basis, conduct DPIAs, and address the EDPB's specific concerns about web scraping's `large volume' and `indiscriminate collection'. The four extraction channels -- platform APIs (clear but restrictive), user-mediated collection (consent-based but limited scale), third-party aggregators (convenient but legally precarious), and self-directed scraping (flexible but highest exposure) -- each present distinct trade-offs between legal certainty and research utility.

\noindent\rule{\columnwidth}{0.4pt}

\noindent\textbf{Decision: Data Acquisition Method}

\noindent\textbf{Selected approach:} Reddit API (official, free tier)

\noindent\textbf{Rationale:}
\begin{itemize}
\item Compliant with platform terms of service
\item 100 queries per minute is sufficient for research needs
\item Provides structured data with consistent formatting
\item Avoids technical confrontation with platform
\end{itemize}

\noindent\rule{\columnwidth}{0.4pt}

\noindent\textbf{Implementation Steps:}
\begin{itemize}
\item Register for Reddit API access with academic credentials
\item Configure API client with rate limiting
\item Document Article 9 legal basis: Political opinions are special category data, but processing justified under Article 9(2)(j) for scientific research with appropriate safeguards
\item Prepare Article 14 public notice for university website explaining research and data subject rights
\item Implement extraction script focusing on data minimisation
\end{itemize}

\noindent\textbf{DPIA Update -- Extract Phase:}

\noindent\textit{Method confirmed:} Reddit API v2, authenticated access

\noindent\textit{Volume extracted:} 100,000 posts across 3 subreddits

\noindent\textit{Special categories:} Political opinions present, Article 9(2)(j) basis documented

\noindent\textit{Notification approach:} Public notice published at university.edu/political-discourse-research

\noindent\textit{Technical measures:} HTTPS only, temporary storage, access logging enabled

\subsection{Load -- Secure Storage of Raw Data}

This phase has established that Recital 26's anonymisation standard -- data that cannot be re-identified by `any means reasonably likely to be used' -- remains practically unachievable for social media's rich behavioural patterns, meaning datasets stay pseudonymised and fully regulated. Storage therefore requires Article 32's security measures (encryption, access controls, audit logging), whilst international transfers demand adequacy decisions or Standard Contractual Clauses under Articles 45-46. DSM Article 3(2) explicitly authorises retention `for scientific research, including verification of results' -- distinct from GDPR's Article 89 research provisions -- though Reddit's Data API Terms impose additional contractual obligations prohibiting third-party sharing without consent. Major cloud providers offer pre-approved Data Processing Addenda, but researchers must verify coverage for their specific architecture.

\noindent\textbf{Methodological Note:} Whilst the PETLP framework presents a sequential pipeline, social media research often necessitates an Extract-Load-Transform (ELT) approach rather than traditional ETL. As noted in Section 3.1, research workflows are rarely linear -- methodological stages overlap or iterate unpredictably. In this case study, we adopt ELT to enable exploratory analysis and preserve raw data for verification under DSM Article 3(2). This approach allows transformation decisions to be informed by actual data characteristics rather than predetermined assumptions, whilst maintaining an immutable baseline for reproducibility. However, immediate loading of raw data heightens security obligations under Article 32 GDPR, as the stored data remains in its most identifiable form until transformation occurs.

\noindent\textbf{Storage Implementation:}
\begin{itemize}
\item Store raw JSON responses on university servers (EU-based)
\item Implement encryption at rest (AES-256)
\item Set access controls: researcher + supervisor only
\item Enable audit logging for all data access
\item Document retention period: 5 years under DSM Article 3(2) for scientific verification
\end{itemize}

\noindent\textbf{DPIA Update -- Load Phase:}

\noindent\textit{Storage location:} EU-based university servers, no international transfers

\noindent\textit{Security measures:} AES-256 encryption at rest, TLS 1.3 for connections

\noindent\textit{Access controls:} Role-based permissions, two authorised users only

\noindent\textit{Retention schedule:} 5-year maximum, annual review requirement

\noindent\textit{Deletion protocols:} Automated alerts at 4.5 years, cascade deletion across backups

\subsection{Transform -- Privacy-Preserving Data Preparation}

We have shown that transformation triggers InfoSoc Article 2's broad reproduction right with every preprocessing operation, though DSM Article 3 permits these for qualifying researchers conducting scientific research, including retention under Article 3(2) for verification. Simultaneously, GDPR Article 5(1)(c) mandates data minimisation -- researchers cannot retain fields ``just in case'' but must justify each element's necessity. The EDPB specifically expects `data minimisation strategies' even at preprocessing stages, whilst warning that achieving true anonymisation proves nearly impossible with high-dimensional social media data (de Montjoye demonstrated 95\% re-identification with just four data points). Differential privacy emerges as state-of-the-art protection, though even aggregated features risk membership inference and model inversion attacks.

\noindent\rule{\columnwidth}{0.4pt}

\noindent\textbf{Decision: Privacy Protection Level}

\noindent\textbf{Selected approach:} Moderate protection with systematic de-identification
\begin{itemize}
\item Remove all usernames and user IDs
\item Generalise timestamps to week-level
\item Hash post IDs to prevent re-identification via Reddit search
\item Remove quoted usernames within post content
\end{itemize}

\noindent\rule{\columnwidth}{0.4pt}

\noindent\textbf{DPIA Update -- Transform Phase:}

\noindent\textit{Data minimisation:} Retained only selftext, subreddit, generalised timestamps

\noindent\textit{Anonymisation attempts:} Pseudonymisation applied (usernames removed, IDs hashed)

\noindent\textit{Differential privacy considered:} Not implemented due to utility trade-offs for classification task

\noindent\textit{Special category safeguards:} Political opinions aggregated at community level where possible

\noindent\textit{Intermediate copies:} Secured with AES-256 encryption, deletion scheduled post-processing

\subsection{Present -- Model Training and Dissemination}

This phase reveals the unique challenges of public dissemination: whilst open science mandates require data sharing, the EDPB warns that AI models `cannot, in all cases, be considered anonymous' as training data may remain `absorbed' in model parameters. The LAION decision protects dataset creation but explicitly not distribution, leaving model publication in legal uncertainty until pending CJEU guidance arrives. Researchers must therefore balance multiple risks -- verbatim quotes enabling search-based re-identification, potential copyright liability from model outputs reproducing training data, and platform terms (Reddit explicitly prohibits AI training) -- whilst satisfying transparency requirements through alternative methods like synthetic datasets or identifier-only releases.

\noindent\rule{\columnwidth}{0.4pt}

\noindent\textbf{Decision: Dissemination Strategy}

\noindent\textbf{Initial plan:} Publish paper + fine-tuned model

\noindent\textbf{Legal barrier:} Reddit prohibits derivative models

\noindent\textbf{Revised approach:}
\begin{enumerate}
\item Contact Reddit for academic use permission
\item If denied: Publish methodology and results only (no model)
\item If approved: Implement additional safeguards
\end{enumerate}

\noindent\rule{\columnwidth}{0.4pt}

\noindent\textbf{DPIA Update -- Present Phase:}

\noindent\textit{Re-identification risks:} No verbatim quotes in publication, only paraphrased examples

\noindent\textit{Model privacy:} Differential privacy not applied (permission pending)

\noindent\textit{Copyright considerations:} Model weights not distributed without platform permission

\noindent\textit{Dissemination method:} Aggregated statistics only, no individual-level data

\noindent\textit{Transparency balance:} Methodology detailed, synthetic examples provided

\subsection{Research Outcome Summary}

\noindent\textbf{Compliance Achievements:}
\begin{itemize}
\item[$\checkmark$] GDPR compliance through systematic de-identification
\item[$\checkmark$] DSM Article 3 research exemption properly invoked
\item[$\checkmark$] DPIA maintained as living document throughout
\item[$\checkmark$] Platform terms respected (API use, permission sought)
\end{itemize}

\noindent\textbf{Research Outputs:}
\begin{itemize}
\item Academic paper with aggregated findings
\item Model release contingent on Reddit permission
\item No raw dataset release (privacy risks)
\item Methodological contribution to privacy-preserving research
\end{itemize}

\newpage
\section{Selected EDPB Paragraphs}

The following sections extract and organise key paragraphs from \textbf{Opinion 28/2024 on certain data protection aspects related to the processing of personal data in the context of AI models} \cite{EDPB2024Opinion28} (hereafter Opinion), \textbf{Report on the work undertaken by the ChatGPT Taskforce} \cite{EDPB2024ChatGPTReport} (hereafter Report), and \textbf{Guidelines on transparency under Regulation 2016/679} \cite{WP292018Transparency} (hereafter Guidelines). Quotations are grouped by common researcher concerns, with practical implications provided after each thematic section.

\subsection{Can Researchers Scrape Public Social Media Data?}

\subsubsection{The Public $\neq$ Consent Principle}
~\\

``Regarding the processing of special categories of personal data... However, \textit{the mere fact that personal data is publicly accessible does not imply that `the data subject has manifestly made such data public'}... \textit{it is important to ascertain whether the data subject had intended, explicitly and by a clear affirmative action, to make the personal data in question accessible to the general public}.'' (\textbf{Report, Paragraph 18})
~\\

``The use of web scraping in the development phase may lead - in the absence of sufficient safeguards - to \textit{significant impacts on individuals, due to the large volume of data collected, the large number of data subjects, and the indiscriminate collection of personal data}.'' (\textbf{Opinion, Paragraph 86})
~\\

\textbf{Practical Implications}: Publicly visible social media posts are not automatically fair game for research. Researchers must consider whether users intended their data to be used for AI training -- posting on social media does not equal consent for all uses. This is particularly critical for sensitive data (health information, political views, etc.) where additional safeguards are required.

\subsection{When Can Researchers Avoid Individual Notification?}

\subsubsection{The Impossibility Standard}
~\\

``The situation where it `proves impossible' under Article 14.5(b) to provide the information is \textit{an all or nothing situation because something is either impossible or it is not; there are no degrees of impossibility}... \textit{In practice, there will be very few situations in which a data controller can demonstrate that it is actually impossible} to provide the information to data subjects.'' (\textbf{Guidelines, Paragraph 59})

\subsubsection{The Disproportionate Effort Exception}
~\\

``Where a data controller seeks to rely on the exception in Article 14.5(b) on the basis that provision of the information would involve a disproportionate effort, \textit{it should carry out a balancing exercise to assess the effort involved... against the impact and effects on the data subject}... This assessment should be documented... In such a case, Article 14.5(b) specifies that \textit{the controller must take appropriate measures to protect the data subject's rights, freedoms and legitimate interests}... One appropriate measure... \textit{that controllers must always take is to make the information publicly available}.'' (\textbf{Guidelines, Paragraph 64})
~\\

``...Given the emphasis in Recital 62 and Article 14.5(b) on archiving, research and statistical purposes... WP29's position is that \textit{this exception should not be routinely relied upon by data controllers who are not processing personal data for the purposes of archiving in the public interest, for scientific or historical research purposes or statistical purposes}.'' (\textbf{Guidelines, Paragraph 61})
~\\

``...Bearing in mind that \textit{the development phases of AI models may involve the collection of large amounts of data from publicly accessible sources (such as via web scraping techniques), reliance on the exception provided under Article 14(5)(b) GDPR is strictly limited} to when the requirements of this provision are fully met.'' (\textbf{Opinion, Paragraph 63})
~\\

\textbf{Practical Implications}: Researchers cannot simply claim it's ``too hard'' to notify millions of social media users. They must: (1) Document why individual notification would require disproportionate effort; (2) Balance their effort against potential harm to data subjects; (3) Always publish a public privacy notice on their institutional website; (4) Note that this exception is more readily available for genuine research purposes than commercial applications.

\subsection{What Legal Basis Should Researchers Use?}

\subsubsection{Legitimate Interest Requirements}
~\\

``Regarding web scraping, OpenAI brought forward Article 6(1)(f) GDPR as legal basis... the legal assessment of Article 6(1)(f) GDPR should be based on the following criteria: i) existence of a legitimate interest, ii) necessity of processing... and iii) balancing of interests. \textit{Fundamental rights and freedoms of data subjects on one hand and the controller's legitimate interests on the other hand have to be evaluated and balanced carefully}.'' (\textbf{Report, Paragraph 16})
~\\

``...the intended volume of personal data involved in the AI model needs to be assessed in light of \textit{less intrusive alternatives that may reasonably be available} to achieve just as effectively the purpose of the legitimate interest pursued. If the pursuit of the purpose is also possible through an AI model that does not entail processing of personal data, then processing personal data should be considered as not necessary.'' (\textbf{Opinion, Paragraph 73})
~\\

``The EDPB recalls that \textit{the GDPR does not establish any hierarchy between the different legal bases} laid down in Article 6(1) GDPR.'' (\textbf{Opinion, Paragraph 60})
~\\

\textbf{Practical Implications}: Researchers must complete and document a three-part test: (1) Identify a clear legitimate interest (e.g., scientific research advancing public knowledge); (2) Prove the data processing is necessary (consider: Could synthetic data work instead?); (3) Balance their interests against data subjects' rights (research purpose may tip the balance, but doesn't guarantee it).

\subsection{How Should Researchers Design Privacy-Preserving AI Models?}

\subsubsection{Data Minimisation in Collection and Processing}
~\\

``The first evaluation area involves examining the selection of sources used to train the AI model... any steps taken to \textit{avoid or limit the collection of personal data}...'' (\textbf{Opinion, Paragraph 50})
~\\

``The second area of evaluation relates to the preparation of data for the training phase... (iii) \textit{the data minimisation strategies and techniques employed to restrict the volume of personal data included in the training process}...'' (\textbf{Opinion, Paragraph 51})
~\\

``...the development and deployment of AI models requires that \textit{personal data should be adequate, relevant and necessary in relation to the purpose}. This can include the processing of personal data to avoid the risks of potential biases and errors when this is clearly and specifically identified within the purpose...'' (\textbf{Opinion, Paragraph 64})
~\\

\textbf{Practical Implications}: Researchers should implement data minimisation at every stage: Collection: Define specific criteria for what data to collect (not ``everything available''); Preprocessing: Filter out irrelevant personal data before training; Only collect personal data when necessary for addressing bias or specific research questions.

\subsubsection{Technical Privacy Safeguards}
~\\

``The third area of evaluation concerns the selection of robust methods in AI model development... (ii) \textit{whether the controller implemented appropriate and effective privacy-preserving techniques (e.g. differential privacy)}.'' (\textbf{Opinion, Paragraph 52})
~\\

``...in line with the principle of accountability... \textit{controllers processing personal data in the context of LLMs shall take all necessary steps to ensure full compliance with the requirements of the GDPR}. In particular, technical impossibility cannot be invoked to justify non-compliance... especially considering that \textit{the principle of data protection by design... shall be taken into account at the time of the determination of the means for processing and at the time of the processing itself}.'' (\textbf{Report, Paragraph 7})
~\\

\textbf{Practical Implications}: ``It's technically difficult'' is not a valid excuse. Researchers must: Implement privacy-preserving techniques like differential privacy from the start; Design systems with privacy in mind, not as an afterthought; Consider privacy at both the algorithm design and implementation stages.

\subsection{Can Researchers Assume Their Models Are Anonymous?}

\subsubsection{The Anonymity Myth}
~\\

``Based on the above considerations, \textit{the EDPB considers that AI models trained on personal data cannot, in all cases, be considered anonymous}. Instead, the determination of whether an AI model is anonymous should be assessed... on a case-by-case basis.'' (\textbf{Opinion, Paragraph 34})
~\\

``The EDPB considers that... \textit{information from the training dataset, including personal data, may still remain `absorbed' in the parameters of the model}... They may differ from the original training data points, but may still retain the original information of those data...'' (\textbf{Opinion, Paragraph 31})
~\\

``...SAs should take into account that \textit{successful testing which covers widely known, state-of-the-art attacks can only be evidence for the resistance to those attacks}... this could include... structured testing against: (i) attribute and membership inference; (ii) exfiltration; (iii) regurgitation of training data; (iv) model inversion; or (v) reconstruction attacks.'' (\textbf{Opinion, Paragraph 55})
~\\

\textbf{Practical Implications}: Never assume a trained model is anonymous just because it doesn't contain raw training data. Researchers must: Test models against known attack vectors (membership inference, data extraction, etc.); Document these tests and their results; Understand that passing current tests doesn't guarantee future anonymity.

\subsubsection{Different Standards for Different Contexts}
~\\

``...the conclusion of a SA's assessment might differ between \textit{a publicly available AI model, which is accessible to an unknown number of people... and an internal AI model only accessible to employees}.'' (\textbf{Opinion, Paragraph 46})
~\\

``...this assessment should be made taking into account \textit{`all the means reasonably likely to be used' by the controller or another person to identify individuals}...'' (\textbf{Opinion, Paragraph 41})
~\\

\textbf{Practical Implications}: The deployment context matters significantly: Models shared publicly face higher anonymisation standards; Internal research models may have lower risk thresholds; Consider who might access your model and what resources they might have for attacks.

\subsection{Documentation and Accountability Requirements}
~\\

``...This also applies to any processing that would include the training of an AI model, even if the objective of the processing is anonymisation. \textit{SAs should consider such documentation... as they are fundamental steps to demonstrate that personal data is not processed}.'' (\textbf{Opinion, Paragraph 56})
~\\

``...In particular... \textit{controllers shall be responsible for, and be able to demonstrate compliance with, all the principles relating to their processing of personal data}.'' (\textbf{Opinion, Paragraph 15})
~\\

``...In light of the complex processing situation and the factual limits for data subjects to intervene, \textit{it is imperative that data subjects can exercise their rights in an easily accessible manner}.'' (\textbf{Report, Paragraph 33})
~\\

\textbf{Practical Implications}: Documentation is not optional bureaucracy -- it's a legal requirement. Researchers must: Document all decisions about data collection, processing, and model design; Maintain records that can demonstrate GDPR compliance; Provide clear, accessible mechanisms for data subjects to exercise their rights (even if most won't use them).

\subsection{Understanding Risks and User Expectations}
~\\

``...For example, \textit{large-scale and indiscriminate data collection by AI models in the development phase may create a sense of surveillance for data subjects}... This may lead individuals to self-censor, and present risks of undermining their freedom of expression...'' (\textbf{Opinion, Paragraph 80})
~\\

``...In the context of the development phase of an AI model, these may include... \textit{the interest in self-determination and retaining control over one's own personal data (e.g. the data gathered for the development of the model)}.'' (\textbf{Opinion, Paragraph 77})
~\\

``...Recital 39 stipulates, amongst other things, that data subjects should be \textit{`made aware of the risks, rules, safeguards and rights in relation to the processing of personal data and how to exercise their rights in relation to such processing'}...'' (\textbf{Guidelines, Paragraph 28})
~\\

\textbf{Practical Implications}: Consider the broader societal impact: Large-scale scraping may cause users to self-censor online; Users have legitimate interests in controlling their data's use in AI training; Transparency about risks and safeguards isn't just good practice -- it's legally required.

\newpage
\section{Selected Excerpts from Reddit's Terms and Policies}

The following sections extract key provisions from Reddit's terms and policies as of 2025: the \textbf{User Agreement} (effective 28 June 2025) \cite{RedditUserAgreement}\footnote{https://redditinc.com/policies/user-agreement}, \textbf{Public Content Policy} (updated 30 May 2025) \cite{RedditContentPolicy}\footnote{https://support.reddithelp.com/hc/en-us/articles/26410290525844-Public-Content-Policy}, \textbf{Developer Terms} (effective 24 September 2024) \cite{RedditDeveloperTerms}\footnote{https://redditinc.com/policies/developer-terms}, \textbf{Data API Terms} (effective 19 June 2023) \cite{RedditDataAPITerms2024}\footnote{https://redditinc.com/policies/data-api-terms}, and \textbf{Developer Platform Wiki} \cite{RedditDataAPIWiki2024} (updated 15 February 2025)\footnote{https://support.reddithelp.com/hc/en-us/articles/14945211791892-Developer-Platform-Accessing-Reddit-Data}. 

\subsection{Platform Position on Public Data and AI}

\subsubsection{Reddit's Stance on Data Scraping}
~\\

``Unfortunately, we see more and more entities using unauthorized access (for example, by scraping or using data brokers) or misusing authorized access to collect public data in bulk, \textit{especially with the rise of use cases like generative AI}. These entities amass public data, including Reddit content, for their own commercial gain, with no perceived limits to their use of that data, and \textit{with no regard for user rights or privacy}.'' (\textbf{Public Content Policy})
~\\

``We still believe in an open internet, but \textit{we do not believe that third parties have a right to misuse public content just because it's public}.'' (\textbf{Public Content Policy})

\subsubsection{Commercial Licensing Model}
~\\

``Reddit may license public content for commercial or non-commercial use: To address this misuse of public Reddit content, we enter into licensing arrangements that allow us to put into place meaningful protections.'' (\textbf{Public Content Policy})
~\\

``Our data licensees are primarily: ... \textit{Large language model makers that share our values that not all public data on the open internet is free to use without restraint}...'' (\textbf{Public Content Policy})
~\\

``Can I use content on Reddit to build a large language / AI model? No. You may not use content on Reddit as an input for any model training without explicit consent from Reddit. \textit{Commercial use of any model trained with Reddit data is prohibited without explicit approval}.'' (\textbf{Developer Platform Wiki})

\subsection{Technical and Legal Restrictions on AI Training}
~\\

``You will not... access or use the Reddit Services and Data through any means (including by accessing our API or indexing, caching, or crawling our Reddit Services and Data) to \textit{train large language, artificial intelligence, or other algorithmic models or related services without our permission}'' (\textbf{Developer Terms, Section 4})
~\\

``Except as expressly permitted by this section, no other rights or licenses are granted or implied, \textit{including any right to use User Content for other purposes, such as for training a machine learning or AI model}, without the express permission of rightsholders in the applicable User Content.'' (\textbf{Data API Terms, Section 2.4})
~\\

\textbf{Practical Implications}: These provisions create multiple layers of restriction: (1) technical restrictions through API terms, (2) intellectual property restrictions recognising user ownership, and (3) contractual prohibitions on AI training without explicit permission.

\subsection{Research Exceptions and Limitations}
~\\

``Use for research purposes is OK provided you use it \textit{exclusively for academic (i.e. non-commercial) purposes}, and don't redistribute our data or any derivative products or services based on our data \textit{(e.g. models trained using Reddit data)}.'' (\textbf{Developer Platform Wiki})
~\\

``You can publish the results of your research, so long as you \textit{exclude our data or any derivative products based on our data (e.g. models trained using Reddit data)}, you credit Reddit, and anonymize information in your published results.'' (\textbf{Developer Platform Wiki})
~\\

\textbf{Practical Implications}: Reddit permits limited academic research but explicitly prohibits: (1) commercial use of research outputs, (2) distribution of trained models, and (3) publication of the underlying data. Researchers must also provide advance notice before publication.

\newpage
\section{Selected Database, InfoSoc and DSM Directive Provisions}

The following sections extract and organise key provisions from \textbf{Directive 96/9/EC on the legal protection of databases} \cite{Directive96_9} (hereafter Database Directive), \textbf{Directive 2001/29/EC on the harmonisation of certain aspects of copyright and related rights in the information society} \cite{InfoSocDirective2001} (hereafter InfoSoc Directive), and \textbf{Directive (EU) 2019/790 on copyright and related rights in the Digital Single Market and amending Directives 96/9/EC and 2001/29/EC} \cite{DSMDirective2019} (hereafter DSM Directive) relevant to researchers training AI models on social media data. Quotations are grouped thematically, with practical implications provided after each section.

\subsection{What Copyright and Database Rights Apply to Social Media Data?}

\subsubsection{Copyright Protection - Reproduction Rights}
~\\

``Member States shall provide for the exclusive right to authorise or prohibit \textit{direct or indirect, temporary or permanent reproduction} by any means and in any form, \textit{in whole or in part}: (a) for authors, of their works'' (\textbf{InfoSoc Directive, Article 2})
~\\

``Temporary acts of reproduction referred to in Article 2, which are \textit{transient or incidental} [and] an integral and essential part of a technological process and whose sole purpose is to enable: (a) a transmission in a network between third parties by an intermediary, or (b) a lawful use of a work or other subject-matter to be made, and which have \textit{no independent economic significance}, shall be exempted from the reproduction right'' (\textbf{InfoSoc Directive, Article 5(1)})
~\\

\subsubsection{Copyright Protection - Communication to the Public}
~\\

``Member States shall provide authors with the exclusive right to authorise or prohibit any \textit{communication to the public} of their works, by wire or wireless means, including the \textit{making available to the public} of their works in such a way that members of the public may access them from a place and at a time individually chosen by them.'' (\textbf{InfoSoc Directive, Article 3(1)})
~\\

\subsubsection{Database Rights Protection}
~\\

``Member States shall provide for a right for the maker of a database which shows that there has been \textit{qualitatively and/or quantitatively a substantial investment}... to prevent extraction and/or re-utilization of \textit{the whole or of a substantial part}... of the contents of that database.'' (\textbf{Database Directive, Article 7(1)})
~\\

``The \textit{repeated and systematic extraction} and/or re-utilization of insubstantial parts of the contents of the database implying acts which \textit{conflict with a normal exploitation} of that database or which unreasonably prejudice the legitimate interests of the maker of the database \textit{shall not be permitted}.'' (\textbf{Database Directive, Article 7(5)})
~\\

\textbf{Practical Implications}: Social media platforms enjoy multiple layers of protection. Individual posts may be protected by copyright if they meet originality thresholds, whilst the platform's aggregate database benefits from sui generis rights. AI training involves reproduction (creating copies for processing), potential communication to the public (if models regenerate training data), and database extraction (systematic collection of posts). Even temporary copies in RAM during processing require authorisation unless they fall within narrow exceptions.

\subsection{Traditional Research Exceptions vs TDM Rights}

\subsubsection{Pre-DSM Research Exceptions}
~\\

``Member States may provide for exceptions or limitations to the rights provided for in Articles 2 and 3 in the following cases: (a) use for the \textit{sole purpose of illustration for teaching or scientific research}, as long as the source, including the author's name, is indicated, unless this turns out to be impossible and \textit{to the extent justified by the non-commercial purpose to be achieved}'' (\textbf{InfoSoc Directive, Article 5(3)(a)})
~\\

``The exceptions and limitations provided for in paragraphs 1, 2, 3 and 4 shall only be applied in \textit{certain special cases} which do not conflict with a normal exploitation of the work or other subject-matter and \textit{do not unreasonably prejudice the legitimate interests of the rightholder}.'' (\textbf{InfoSoc Directive, Article 5(5)})
~\\

``When applying the exception or limitation for non-commercial educational and scientific research purposes, including distance learning, \textit{the non-commercial nature of the activity in question should be determined by that activity as such}. The organisational structure and the means of funding of the establishment concerned are \textit{not the decisive factors} in this respect.'' (\textbf{InfoSoc Directive, Recital 42})
~\\

\subsubsection{DSM Text and Data Mining Rights}
~\\

``Member States shall provide for an \textit{exception} to the rights provided for in Article 5(a) and Article 7(1) of Directive 96/9/EC, \textit{Article 2 of Directive 2001/29/EC}... for reproductions and extractions made by \textit{research organisations and cultural heritage institutions} in order to carry out, \textit{for the purposes of scientific research}, text and data mining of works or other subject matter to which they have \textit{lawful access}.'' (\textbf{DSM Directive, Article 3(1)})
~\\

``Text and data mining makes the processing of \textit{large amounts of information} with a view to \textit{gaining new knowledge and discovering new trends} possible.'' (\textbf{DSM Directive, Recital 8})
~\\

\textbf{Practical Implications}: The InfoSoc research exception was too narrow for AI training -- limited to ``illustration'' and subject to the three-step test. The DSM Directive amends this by creating specific TDM rights that override both copyright (InfoSoc Article 2) and database rights, explicitly permitting large-scale data processing for pattern discovery. However, this broader right remains limited to qualifying research organisations.

\subsection{Can Researchers Extract Social Media Data for AI Training?}

\subsubsection{Database Rights Restrictions}
~\\

``Member States shall provide for a right for the maker of a database which shows that there has been \textit{qualitatively and/or quantitatively a substantial investment}... to prevent extraction and/or re-utilization of \textit{the whole or of a substantial part}... of the contents of that database.'' (\textbf{Database Directive, Article 7(1)})
~\\

``The \textit{repeated and systematic extraction} and/or re-utilization of insubstantial parts of the contents of the database implying acts which \textit{conflict with a normal exploitation} of that database or which unreasonably prejudice the legitimate interests of the maker of the database \textit{shall not be permitted}.'' (\textbf{Database Directive, Article 7(5)})
~\\

\subsubsection{DSM Text and Data Mining Rights}
~\\

``Member States shall provide for an \textit{exception} to the rights provided for in Article 5(a) and Article 7(1) of Directive 96/9/EC... for reproductions and extractions made by \textit{research organisations and cultural heritage institutions} in order to carry out, \textit{for the purposes of scientific research}, text and data mining of works or other subject matter to which they have \textit{lawful access}.'' (\textbf{DSM Directive, Article 3(1)})
~\\

``Text and data mining makes the processing of \textit{large amounts of information} with a view to \textit{gaining new knowledge and discovering new trends} possible.'' (\textbf{DSM Directive, Recital 8})
~\\

\textbf{Practical Implications}: The DSM Directive overrides database rights for academic researchers. Whilst platforms retain database rights that would normally prohibit AI training datasets, Article 3 creates a mandatory exception for scientific research by qualifying institutions. This supersedes both the substantial extraction prohibition and the systematic use restriction when done for non-commercial research purposes.

\subsection{What Constitutes a Research Organisation?}

\subsubsection{Institutional Requirements}
~\\

```research organisation' means a university, including its libraries, a research institute or any other entity, \textit{the primary goal of which is to conduct scientific research} or to carry out educational activities involving also the conduct of scientific research: (a) on a \textit{not-for-profit basis} or by reinvesting all the profits in its scientific research; or (b) pursuant to a \textit{public interest mission} recognised by a Member State; in such a way that the access to the results generated by such scientific research \textit{cannot be enjoyed on a preferential basis by an undertaking that exercises a decisive influence} upon such organisation'' (\textbf{DSM Directive, Article 2(1)})
~\\

``Due to the diversity of such entities, it is important to have a common understanding of research organisations... Conversely, \textit{organisations upon which commercial undertakings have a decisive influence} allowing such undertakings to exercise control... which could result in \textit{preferential access to the results of the research}, should not be considered research organisations for the purposes of this Directive.'' (\textbf{DSM Directive, Recital 12})
~\\

\textbf{Practical Implications}: University researchers clearly qualify, but industry partnerships require careful structuring. Research must be the primary goal, and commercial entities cannot have `decisive influence' or `preferential access' to results. This may affect researchers with significant industry funding or those planning commercial spin-offs.

\subsection{Can Platforms Impose Technical Restrictions?}

\subsubsection{Security Measures Allowed}
~\\

``Rightholders shall be allowed to apply measures to ensure the \textit{security and integrity} of the networks and databases where the works or other subject matter are hosted. Such measures \textit{shall not go beyond what is necessary} to achieve that objective.'' (\textbf{DSM Directive, Article 3(3)})
~\\

``Such measures could, for example, be used to ensure that \textit{only persons having lawful access} to their data can access them, including through IP address validation or user authentication. Those measures should remain \textit{proportionate to the risks involved}... and should \textit{not undermine the effective application of the exception}.'' (\textbf{DSM Directive, Recital 16})
~\\

\subsubsection{Storage and Verification Rights}
~\\

``Copies of works or other subject matter made in compliance with paragraph 1 \textit{shall be stored with an appropriate level of security} and \textit{may be retained} for the purposes of scientific research, including for the \textit{verification of research results}.'' (\textbf{DSM Directive, Article 3(2)})
~\\

\textbf{Practical Implications}: Platforms can implement reasonable security measures but cannot use them to defeat the research exception. Rate limiting or authentication requirements may be permissible if proportionate, but blanket bans on automated access for researchers would likely exceed necessity. Researchers have explicit rights to store data for verification. However, this does not mean that the data can be distributed. 

\subsection{Commercial vs Non-Commercial Research}

\subsubsection{Non-Commercial Requirement}
~\\

``Member States may stipulate that lawful users of a database which is made available to the public in whatever manner may, without the authorization of its maker, extract or re-utilize a substantial part of its contents... (b) in the case of extraction for the purposes of illustration for teaching or scientific research, as long as the source is indicated and \textit{to the extent justified by the non-commercial purpose to be achieved}'' (\textbf{Database Directive, Article 9(b)})
~\\

\subsubsection{Broader Commercial TDM Exception}
~\\

``Member States shall provide for an exception or limitation to the rights... for reproductions and extractions of \textit{lawfully accessible works and other subject matter for the purposes of text and data mining}. Reproductions and extractions made pursuant to paragraph 1 may be retained for as long as is necessary for the purposes of text and data mining. The exception or limitation provided for in paragraph 1 shall apply on condition that the use of works and other subject matter referred to in that paragraph \textit{has not been expressly reserved by their rightholders in an appropriate manner, such as machine-readable means in the case of content made publicly available online}'' (\textbf{Directive 2019/790, Article 4(1)-(3)})
~\\

``In the case of content that has been made publicly available online, it should only be considered appropriate to reserve those rights by the use of \textit{machine-readable means}, including metadata and terms and conditions of a website or a service.'' (\textbf{DSM Directive, Recital 18})
~\\

\textbf{Practical Implications}: Academic research under Article 3 enjoys absolute protection, whilst commercial research under Article 4 can be blocked by platforms through machine-readable reservations. Researchers planning any commercialisation should consider which exception applies, as the transition from academic to commercial use could eliminate their TDM rights.

\subsection{Overall Assessment for AI Researchers}

The combined framework creates a complex multi-dimensional system where researchers' rights depend on institutional status, research purpose, and legal basis:
~\\

\textbf{Research Organisations with Scientific Purpose (Article 3 DSM):}
\begin{itemize}
\item Can extract large volume of social media data for AI training
\item Cannot be stopped by contracts or terms of service
\item Must meet institutional criteria: primary research goal, not-for-profit \textit{or} public interest mission
\item No preferential access for commercial entities with ``decisive influence''
\item Covers both natural and human sciences
\item Can store data securely for verification
\end{itemize}
~\\

\textbf{Public Interest Research by Other Entities:}
\begin{itemize}
\item Commercial entities conducting public interest research do \textit{not} automatically qualify under Article 3
\item Must rely on Article 4 exception (if rightholders haven't reserved rights)
\item Public interest alone insufficient without meeting research organisation criteria
\item May strengthen case under the GDPR legitimate interest balancing test
\end{itemize}
~\\

\textbf{Legitimate Interest-Based Research (GDPR Article 6(1)(f)):}
\begin{itemize}
\item Provides lawful basis for data processing but \textit{not} copyright/database exceptions
\item Must still comply with Database Rights Directive restrictions
\item Academic researchers cannot rely solely on legitimate interest to override TDM restrictions
\item Requires case-by-case balancing against data subjects' rights
\item Does not create immunity from platform terms or technical measures
\end{itemize}
~\\

\textbf{General Commercial/Other Research (Article 4 DSM):}
\begin{itemize}
\item Can only extract if platform hasn't reserved rights via machine-readable means
\item Most major platforms likely to implement opt-outs
\item Database rights remain fully enforceable
\item Need explicit licensing for substantial extraction
\item Includes academic researchers planning commercialisation
\end{itemize}
~\\

The DSM framework prioritises institutional form over research merit. Commercial entities, even if engaged in research with societal benefit, may not enjoy the same exceptions solely by virtue of their research merit or public importance. Legitimate interest under the GDPR provides a lawful basis for processing personal data but offers no relief from copyright or database restrictions. Researchers must assess which framework applies based on their institutional status and research purpose, not just their research's societal value.

\section{Temporal Limitations of Reddit API Data Retrieval}

\begin{figure}[!htbp]
  \centering
  \includegraphics[width=\columnwidth]{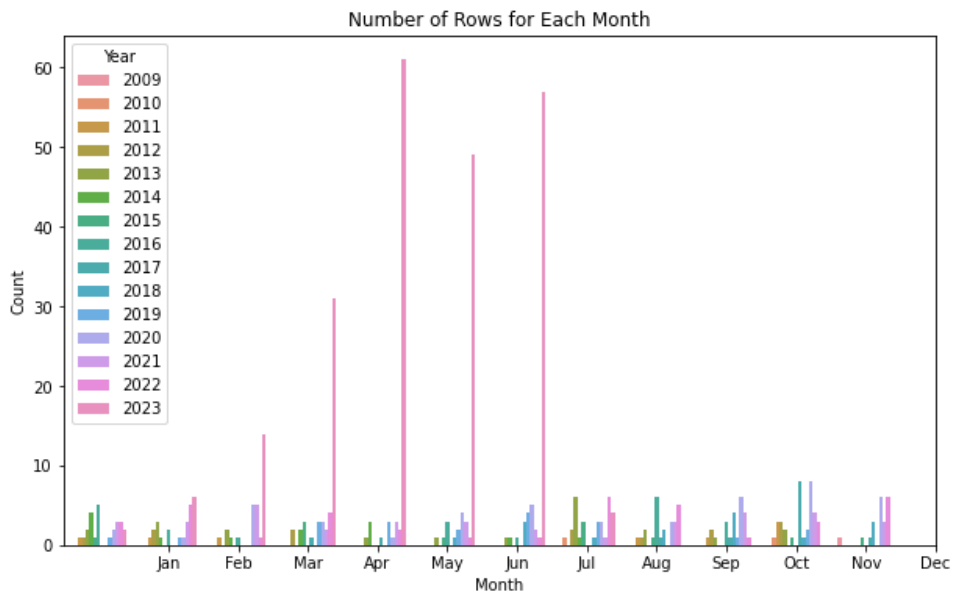}
  \caption{Temporal Distribution of Accessible Reddit Posts (2009-2023).}
  \label{fig:temporal}
\end{figure}

As illustrated in the Figure \ref{fig:temporal}, which displays data collected in June 2023 from 18 international current affairs subreddits (e.g., r/worldnews) using \texttt{PRAW} (Python Reddit API Wrapper), there's a clear limitation in retrieving posts older than approximately six months. The graph shows a consistent pattern across years, with the number of accessible posts dropping dramatically for months prior to January of the current year. This pattern suggests that the Reddit API imposes a rolling six-month window for data retrieval, beyond which historical data becomes increasingly sparse or inaccessible. Notably, this limitation coincides with Reddit's policy of archiving posts that are older than 6 months, which prevents further interactions such as commenting or voting on these older posts. This limitation has important implications for longitudinal studies or research requiring extensive historical data. Researchers utilising Reddit's API should be aware of this constraint and plan their data collection strategies accordingly.

\end{document}